\documentclass[alpha-refs]{wiley-article}
\pdfoutput=1
\usepackage{siunitx}
\usepackage{lineno}
\usepackage{float}
\usepackage{setspace}
\usepackage{textcomp}
\DeclareUnicodeCharacter{2009}{\,}

\papertype{Original Article}

\title{Post-processing output from ensembles with and without parametrised convection, to create accurate, blended, high-fidelity rainfall forecasts}

\author[1\authfn{1}]{Est\'ibaliz Gasc\'on}
\author[1]{Andrea Montani}
\author[1]{Tim D. Hewson}

\contrib[\authfn{1}]{Corresponding author.}

\affil[1]{European Centre for Medium-Range Weather Forecasts, Reading, United Kingdom}

\corraddress{Est\'ibaliz Gasc\'on, European Centre for Medium-Range Weather Forecasts, Reading, United Kingdom.}
\corremail{estibaliz.gascon@ecmwf.int}

\fundinginfo{The authors gratefully acknowledge financial support from the European
Union Connecting~ Europe Facility (CEF) -- Telecommunication Sector
Programme for the MISTRAL -- Meteo~ Italian SupercompuTing poRtAL
project (Grant Agreement No. INEA/CEF/ICT/A2017/1567101).}

\runningauthor{Gasc\'on et al.}

\begin{document}

\maketitle

\begin{abstract}
Flash flooding is a significant societal problem, but related precipitation forecasts are often poor. To address, one can try to use output from convection-parametrising (global) ensembles, post-processed to forecast at point-scale, or con\-vec\-tion-resolving 
limited area ensembles. The new methodology described here combines both. We apply ``ecPoint-rainfall'' post-processing to the ECMWF global ensemble. Alongside we use 2.2km COSMO LAM ensemble output (centred on Italy), and also post-process that, using a scale-selective neighbourhood approach to compensate for insufficient members and to preserve consistently forecast local details. The two resulting scale-compatible components then undergo lead-time-weighted blending, to create the final probabilistic 6h rainfall forecasts. Product creation for forecasters, in this way, constituted the ``Italy Flash Flood use case'' within the EU-funded MISTRAL project; real-time delivery of open access products is ongoing.

One year of verification shows that, of the five components (2 x raw, 2 x post-processed and blended), ecPoint is the most skilful. The post-processed COSMO ensemble adds most value to summer convective events in the evening, when the global model has an underprediction bias. In two typical heavy rainfall case studies we observed underestimation of the largest point totals in the raw ECMWF ensemble, and overestimation in the raw COSMO ensemble. However, ecPoint elevated the ECMWF maxima and highlighted best the most affected areas, whilst post-processing of COSMO diminished extremes by eradicating unreliable detail, such that the final merged products looked best from a user perspective, and seemed to be the most skilful of all. Although our LAM post-processing does not implicitly include bias-correction (a topic for further work) our study nonetheless provides a unique blueprint for successfully combining ensemble rainfall forecasts from global and LAM systems around the world. It also has important implications for forecast products as global ensembles move ever closer to having convection-permitting resolution.

\end{abstract}%

\doublespacing
\section{~Introduction}

Short-duration convective precipitation extremes are generally
responsible for flash flooding events, which are a significant societal problem. This paper looks at new ways to anticipate such events by applying novel post-processing techniques to the output of different types of numerical model. 

Many times, numerical weather preciction (NWP) systems cannot predict precipitation with sufficient accuracy in terms of
rainfall quantity and the particular location of the rainfall
events \citep{Silvestro_2015, Gascon_2016}. This is partially because weather forecasts
are often required for points and not, for example, for the large areas of
several hundred square kilometres represented by current-generation global model grid boxes. This
discrepancy can be addressed using high-resolution limited-area models
(LAMs). Alternatively one can apply some post-processing methods to global
forecast models, to convert from gridbox to
point-forecasts \citep{Gneiting_2014, Pillosu_Hewson2021}. 
To better understand these model limitations, we should consider the concept of
sub-grid variability, previously discussed by \citet{Pillosu_Hewson2021}.
Sub-grid variability denotes the variation seen amongst point values within a given model gridbox. In very localised convective
precipitation, which is related to flash floods occurrence, the sub-grid
variability is often so high that forecast totals from globel models of current day resolution will 'miss' the event, by maybe an order of magnitude. However, in large-scale precipitation, the sub-grid
variability in the gridbox is usually considerably less. So a major challenge here for post-processing is dealing with variations in sub-grid variability. Meanwhile, although LAMs, with their much higher resolution, produce more realistic features
than global models, small-scale predictability, if it exists, can only
be maintained for a number of hours, because errors often grow rapidly \citep{Melhauser_2012, Radhakrishna_2012}, particularly in convective situations.

Another reason for the inaccuracy of heavy rainfall forecasts is the innate forecast uncertainties that affect all NWP, which can in principle be represented and quantified using ensemble prediction systems \citep{Buizza_2018}.
Ensembles have been used for several decades to predict the possible evolutions of
large-scale patterns beyond the short-range. In recent years, their use has also become more and more widespread in the framework of LAMs, that have 
convection-permitting resolution (say <5 km in the horizontal), to
facilitate also probabilistic prediction of small scale details that the global models innately cannot capture. The number of LAM ensemble members that are strictly needed to address all the convective degrees of freedom in the atmosphere exceeds what is computationally affordable (usually about 20 members) by at least one order of magnitude even on day 1 \citep{Necker_2020, Kobayashi_2020}, and after that, as synoptic-scale spread among members grows, even more would be needed. Indeed in 2017, \citet{Raynaud_2017} explored how forecast skill depends on ensemble size for LAMs, which tend to have fewer members than global ensembles. Their study compared the benefits of increasing  ensemble size from 12 to 34 with the benefits of increasing horizontal resolution from 2.5 to 1.3 km and concluded that for lead times greater than 12h, increasing ensemble size provides more benefit. 

The COSMO (COnsortium for Small-scale MOdelling) ensemble over Italy is an example of a LAM ensemble. This is a system recently developed for Italy, that runs on an operational-experimental basis, with 2.2 km horizontal resolution, 65
vertical levels and 20 members (COSMO-2I-EPS, or COSMO raw, henceforth). The main features of the system, reported under \url{http://cosmo-model.org/content/tasks/operational/default.htm}, will be further discussed in the next sections. Researchers have also looked at alternative cheap approaches to solve the problem related to the lack of members, such as neighbourhood techniques and lag-based ensembles, to increase ensemble size without increasing computing resources \citep{Ebert_2008, Bouallegue_2013b, DelSole_2017, Schwartz_2017}. The  ``neighbourhood'' post-processing technique has as its
central premise the notion of limited small-scale predictability. It acknowledges that
it is unrealistic to expect high-resolution models to be accurate at the
grid-scale. Several approaches, including the fractional coverage \citep{Theis_2005} and maximum neighborhood value methods \citep{Hitchens_2013}, have been developed to in effect increase the number of members, and thereby increase also the spread, to provide a more comprehensive description of the spatial uncertainties than the raw LAM ensemble can provide on its own. A related challenge is how to use and evaluate the ensembles in convective situations, where the ensemble mean is unlikely to be physically
appropriate for anticipating precipitation extremes \citep{Ancell_2013, Dey_2016b}.

Studies have demonstrated that the skill of convective scale forecasts is very scale-dependent, and this also holds true for ensemble forecasts, wherein ensemble skill increases with both spatial scale and ensemble
size \citep{Roberts_2008, Clark_2011, Bouallegue_2013, Mittermaier_2014}.

So, notwithstanding the above limitations, it
should still be possible to give useful probabilistic forecasts of local precipitation by anticipating uncertainties in the precipitation
location (and assuming that the ensemble is unbiased) \citep{Dey_2016b}. One strategy to address localised precipitation uncertainties
is the ``scale-selective neighbourhood approach''.
Following \citet{Dey_2016a,Dey_2016b} and \citet{Blake_2018}, a similarity 
criterion among all possible ensemble forecast pairs is developed.
The rationale is that we preserve \emph{only the most reliable} fine-scale rainfall signals by (i) retaining that detail in the rainfall patterns wherever ensemble rainfall forecasts are similar to one another (e.g. where precipitation is strongly forced by orographic uplift) whilst (ii) ironing out the detail where there is less agreement (e.g. 'random' summer convection over relatively flat terrain). For each model grid point, the scale-selective neighbourhood approach aims first to quantify the ``agreement scale'' among the ensemble members. This is a metric that describes the spatial scale at which ensemble forecast can be deemed sufficiently similar to each other, according to a criteria described in the next sections. One can think of the agreement scale as being proportional to the length scale across which the post-processing combines together the raw forecast data; i.e. agreement scale = 2 means that data from two extra gridboxes in all directions around the gridbox in question is used, and so forth. So by construction, the smaller the agreement scale is (its minimum value is 0), the higher the agreement among members is.

For global model output, as noted above, an option to address sub-grid variability issues with convective precipitation is to convert gridbox forecasts to point forecasts by using a calibrated post-processing technique. The most common forecast post-processing techniques tend to have limitations associated, such as the large
training datasets needed \citep{Wilks_2007,Mendoza_2015, Gneiting_2014, Hamill_2017, Scheuerer_2015},
the location dependance of the calibration method \citep{Hamill_2018, Taillardat_2020} or the difficulty 
of improving the forecasts for extremes \citep{Scheuerer_2015, Whan_2018, Friederichs_2018, Taillardat_2019}. ``ecPoint-Rainfall'' is a new
probabilistic non-parametric methodology developed at the European Centre
for Medium-Range Weather Forecasts (ECMWF), which successfully addresses these challenges. It currently delivers, on the ECMWF ecCharts platform, experimental real-time probabilistic 12-h precipitation forecasts based on output of the ensemble of the ECMWF Integrated Forecasting System (IFS ENS) \citep{Pillosu_Hewson2017, Pillosu_Hewson2021}. Its main premise is that the forecast-versus-point-observation relationship depends not on location, but on physical considerations, allowing one to foresee different degrees of sub-grid
variability and grid-scale bias according to the prevailing (gridbox) weather situation. It associates each gridbox from a given ensemble member at a given time with a specific "weather type". The output from the ecPoint calibration process is summarised in mapping functions of forecast errors typical for each weather type. These forecast errors are calculated assuming that there is a typical distribution of point rainfall outcomes in a gridbox (as would be measured by multiple evenly-spaced raingauges) for each specific weather type. This distribution will naturally account for the sub-grid variability, but innately also, via averaging, for the grid-scale NWP bias typical for that weather type. This
"calibration by weather types" approach allows us to overcome the achilles heel (from an observation availability perspective) of location-dependence, and indeed create a suitably large calibration dataset for the whole world using only 1-year of observations and 48-hour model (Control run) forecasts. 

The upshot of using the ecPoint approach is that there are clear benefits for flash flood forecasting because extremes are forecast much better (e.g. more than 50 mm rainfall in 12-h). At the same time predictions of smaller totals are also markedly improved \citep{Pillosu_Hewson2021}. Ultimately, use of ecPoint ensures that the horizontal scales of global-ensemble-related output are much more compatible with the scales of the LAM ensemble (2.2 km) meaning that a probabilistic high-resolution blending activity can then be applied (pursuant to blending requirements cited by \cite{Vannitsem_2021}). Here we assume that because a current-day km-scale LAM is able to explicitly represent at least major convective updraughts, it should be reasonably representative of point measurements (in both stratiform and convective situations). For striking evidence of raingauge-relative improvements in multi-threshold frequency biases that can be delivered by even modest model resolution enhancement, most notably in summer, see plots in \cite{Batignani_2019} (p11 and p13, for 9km and 5km resolution models respectively).

Ours is in fact quite a unique approach. In the literature and in operational systems, rainfall blending is much more commonly applied in nowcasting, that just extends a few hours ahead, using for example radar-based predictions alongside NWP (e.g. \cite{Lin_2020}). \cite{Hamill_2017} describe one multi-ensemble initiative for longer leads, although that provides probabilistic output only for a single class (>0.25mm/12h).

ECMWF participated in the MISTRAL (Meteo Italian SupercompuTing PoRtAL) project (http://www.mistralportal.eu) with the primary goal of improving probabilistic rainfall forecast products, to help with the prediction of flash floods in Italy and nearby Mediterranean regions. To do that, the scale-selective neighbourhood technique developed by \citet{Dey_2016b} was adapted and applied to the 2.2 km LAM ensemble forecasts COSMO raw. Then, ecPoint-Rainfall for 6-h total precipitation forecasts was also developed during the project, and computed for the global ECMWF IFS ENS following the technique described by \citet{Pillosu_Hewson2021}. Finally, the two post-processed outputs were blended together, for Italy and surrounding areas, to try to fully exploit the most skilful aspects of the two ensemble systems. The blending weights used were simply tapered lead-time dependant, with more weight given to ecPoint at longer leads as LAM details become less reliable, and also to achieve seamlessness when LAM output stops. The overall aim is naturally to create accurate, high fidelity forecasts, but in particular to support the compilation of reliable flash floods warnings as far in advance as possible. The MISTRAL project is funded under the Connecting Europe Facility (CEF) - Telecommunication
Sector Programme of the European Union. The overarching goal of the project was to facilitate and foster the re-use of datasets by weather-dependent communities, to provide added value services using High-Performance Computing (HPC) resources.

\begin{table}[]
\centering
\resizebox{\textwidth}{!}{%
\begin{tabular}{llllll}
\hline
\textbf{Shortname} & \textbf{IFS ENS raw} & \textbf{6-h ecPoint-Rainfall}                                                                  & \textbf{COSMO raw}    & \textbf{COSMO post}                                                      & \textbf{merged product}                                                                   \\ \hline
\textbf{Forecast system}              & ECMWF ensemble raw   & \begin{tabular}[c]{@{}l@{}}6-h Point rainfall forecast \\ based on ECMWF ensemble\end{tabular} & COSMO ensemble for Italy & \begin{tabular}[c]{@{}l@{}}COSMO ensemble\\  post-processed\end{tabular} & \begin{tabular}[c]{@{}l@{}}COSMO ensemble post-processed\\  + 6-h point rainfall\end{tabular} \\
\textbf{Ensemble members/percentiles} & 51 members           & 99 percentiles                                                                                 & 20 members               & 99 percentiles                                                           & 99 percentiles                                                                                \\
\textbf{horizontal resolution}        & $\sim$18 km          & $\sim$18 km                                                                                    & 2.2 km                   & 2.2 km                                                                   & 2.2 km                                                                                        \\ \hline
\end{tabular}%
}
\caption{{Table with the 5 forecast systems evaluated in this paper and their main features}}
\label{forecastsystems}
\end{table}

The paper is divided up as follows: the methods and datasets used in the creation and validation of these new post-processing products are explained in section 2. Section 3 describes the verification procedures (see Table {\ref{forecastsystems}}) and results, and two case studies are also analysed. Finally, concluding remarks are provided in section 4.

\section{DATA and METHODOLOGY}

\subsection{Observations database}

\begin{figure}[H] \centering \includegraphics[width=33pc]{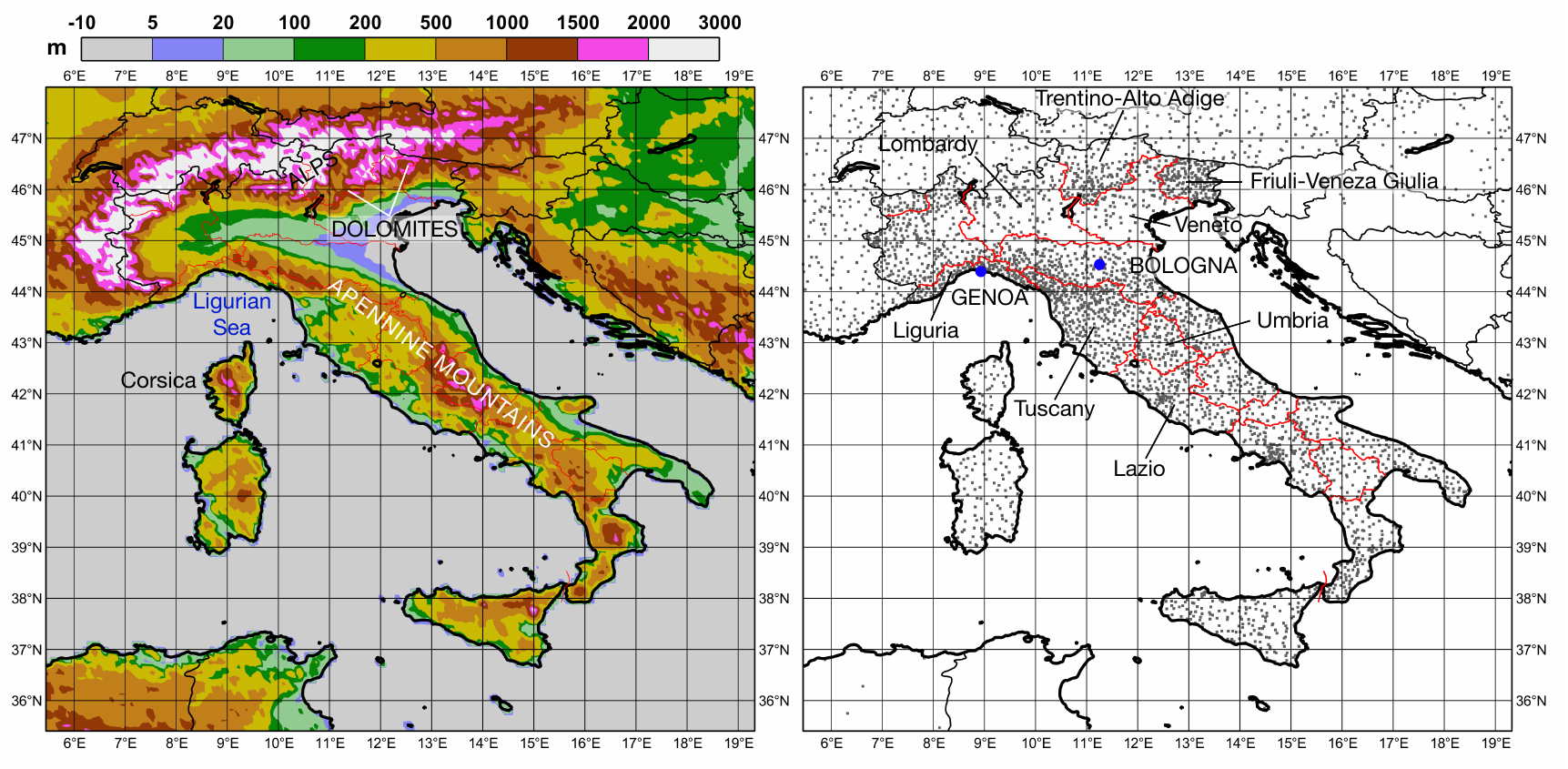} \caption{Maps showing the full domain of the COSMO-2I model, with terrain height in that model (left), and a sample, for one time period, of the SYNOP and HDOBS rain gauge sites used in verification for this paper (right). Labels show the locations of Italian regions (red boundaries), cities and geographical features referenced in the text.}  \label{fig1}
\end{figure}

For verification purposes, ECMWF collects non-real-time high-density observations (HDOBS) from its Member and Cooperating states \citep{Haiden_Duffy2016}, typically received at the end of each month. For Italy these are precipitation-only observations - there about 3500 such sites in the country. The time resolution of this Italian data goes from 6-h to 24-h totals, but only 6-h data has been used in this paper as the main aim was to target and capture short duration extreme rainfall events. In turn this is because for flash flooding incidents the rainfall responsible is typically of this type. The 6-h precipitation data from the HDOBS and from standard SYNOP reports is used to verify our new forecast products, using a standard "nearest neighbour" technique to match station location with model grid box (instead of interpolating between model gridboxes). Figure {\ref{fig1}} shows the topography of the study area from the COSMO raw domain area (left) and the array of rain gauge observations used for verification, in Italy and surrounding countries (right). Note that observation density is equally comprehensive in mountainous areas.

\subsection{ECMWF ensemble post-processing (ecPoint)}

As mentioned before, the new Point Rainfall product aims to deliver probabilities for point measurements of rainfall (within a forecast model gridbox). ``ecPoint'' \citep{Pillosu_Hewson2017, Pillosu_Hewson2021} is the name given to the post-processing philosophy, whilst the companion calibration software
is called ``ecPoint-calibrate''. "ecPoint-Rainfall 12-h" is then a name given to one output type, referring to the fact that its output is for 12-h totals.

ecPoint is based on the concept of conditional verification, and it consists of identifying (through calibration) and correcting (through post-processing) errors in model rainfall forecasts according to a diagnosed gridbox weather-type. Weather-types regularly vary between adjacent gridboxes, between consecutive lead times, and between ensemble members. Each weather type used is defined by a set of value ranges for each one of the different predictors used (i.e. "governing variables"). We typically use 5 to 10 such predictors, and typically create 100-500 weather types. Then all weather types together can be represented by a single decision tree, wherein each level corresponds to a governing variable, each leaf denotes a weather type, and all leaves cover all possible value combinations for all governing variables. A "mapping function" is assigned to each leaf (i.e. to each weather type). Such a mapping function defines the range of (rainfall) forecast errors experienced across the world whenever the said weather type occurred, in the calibration period, in the short range Control run forecasts used for calibration. It ultimately represents the sub-grid variability and the gridscale bias typically seen for those particular (weather) conditions. The final ecPoint-Rainfall product, for a given time interval, is created by accruing together the post-processed forecasts for each ENS member. It can be depicted in percentile or probability format, with the user able to define, according to purpose, the percentile value (e.g. 99th) or the event for the probability (e.g. >50mm/12h).

Following the same methodology as \citet{Pillosu_Hewson2021}, and as part of MISTRAL, a global 6-h
ecPoint-Rainfall product was developed, tested and verified, based on raw IFS ENS forecasts, and using
observations from around the world. The raw IFS ENS comprises 50
perturbed ensemble members and 1 control member, all with a horizontal
resolution of ~18 km. The ecPoint outputs are made available for
overlapping 6-h periods up to day 10, namely T+0-T+6 h,
T+3-T+9 h,..T+234-T+240 h. The main difference compared to the
ecPoint-Rainfall 12-h product arose during calibration, wherein the governing variables and the weather type definitions
were newly customised. ``Weather type'' here refers to specific weather conditions characterising a gridbox, defined by
specific threshold ranges of the governing parameters (e.g. > 75\% of rainfall is denoted to be convective rain; mean 700hPa wind speed during period is in the range 5-20 m/s; ...). Similar to the calibration of
ecPoint-Rainfall 12-h, control run forecast rainfall totals (G) at short-range (3 to 30 h lead times), covering one year (the training
period), were used for calibration by comparing with rainfall observations (r) around the world co-located in space and time. As in \citet{Pillosu_Hewson2021}, cases with G<1 mm (in this case in a 6-h period) were discarded before calibration, for stability and discretisation
reasons, and also we did not use the first three forecast hours either to avoid
possible model rainfall spin-up issues. We merged together two sets of gauge rainfall observations (r): those sourced
through international data-sharing agreements, and others obtained from special
datasets \citep{Haiden_Duffy2016, Haiden2018}. The
calibration period itself was 1 April 2017 to 31 March 2018. The full calibration procedure is not described here but involves
the segregation of all (G,r) pairs into sets, each of which denotes a different gridbox-weather-type. A key aim of segregation was ensuring that each such type has associated with it sub-grid variability levels and/or bias correction factors that differ in a substantive way from those of neighbouring types within the tree. In practice each gridbox-weather-type connects to a mapping function, which describes, in Probability Density Function (PDF) form, the (G,r) relationship for all occasions when the said type occurred during the calibration period. The PDF is recorded using a non-dimensional metric of "Forecast Error Ratio" (FER), defined as follows:

\begin{equation}
\label{eq1}
FER = \frac{(r - G)}{G}
\end{equation}

Later on the mapping function PDFs are used to post-process forecast rainfall totals, according to whenever and wherever the associated weather types re-appear in those forecasts. 

\begin{figure}[H] \centering \includegraphics[width=35pc]{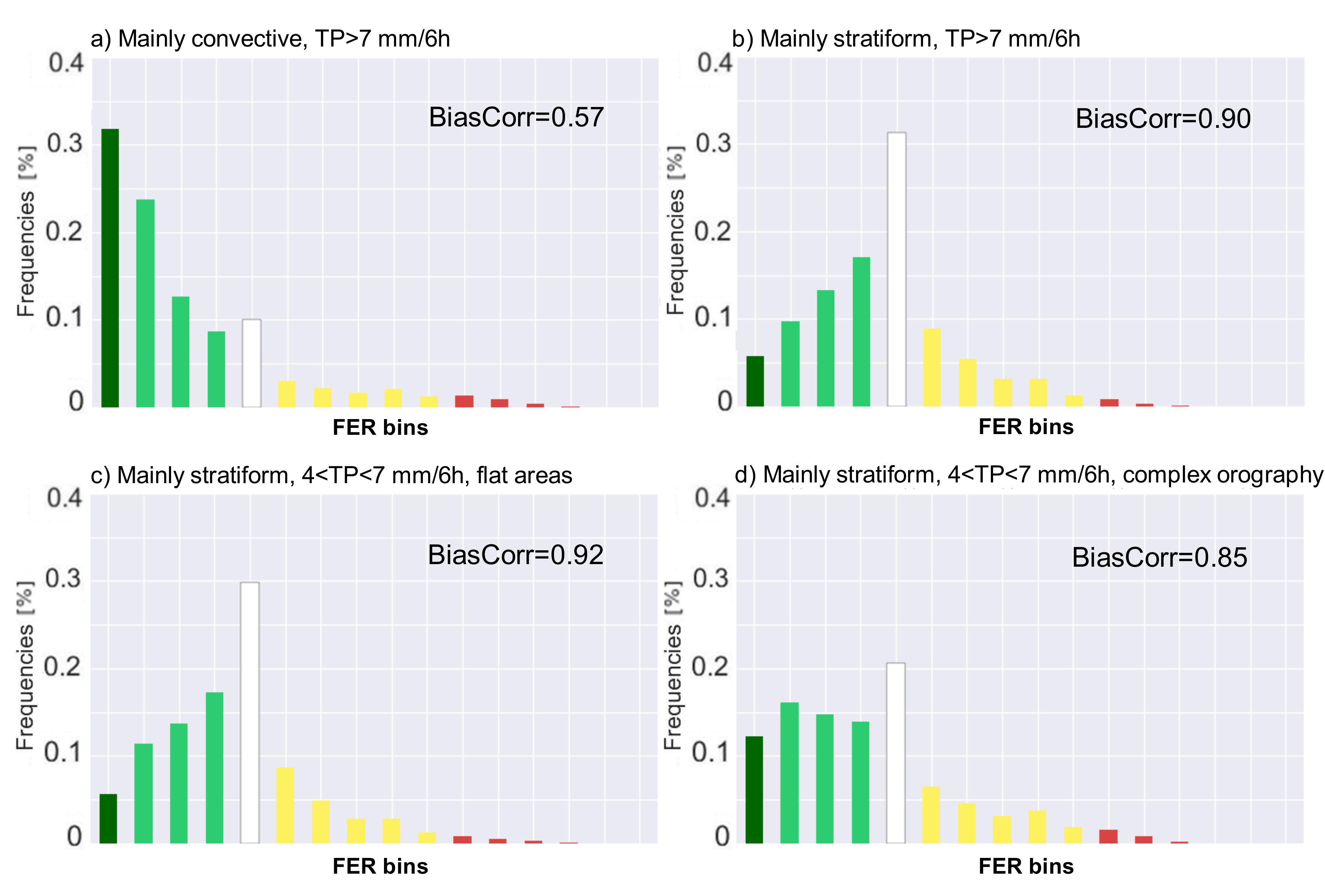} \caption{Mapping function examples for specific weather conditions, shown as
histograms. On (a-d) dark green, green, white, yellow and red bars
denote, respectively, FER ranges for ``mostly dry'' (< -0.99),
``over-prediction'' (-0.99 to -0.25), ``good forecasts'' (-0.25 to
0.25), ``under-prediction'' (0.25 to 2) and ``substantial
under-prediction'' (2 to~\(\infty\)). (a) is for mainly
convective rainfall with large totals forecast, (b) for mainly
stratiform rainfall with large totals forecast, (c) for mainly stratiform with
moderate totals forecast over flat areas and (d) mainly stratiform with
moderate totals forecast in "mountainous" areas. Flat and
mountainous areas were defined using the anciliary model variable "standard deviation of
filtered subgrid orography" (SDfor). ``BiasCorr'' values correspond to the gridbox mean biases (FER) for each weather type.}  \label{fig2}. 
\end{figure}

Some example mapping functions are shown in
Figure \ref{fig2}. Green bars represent
``over-prediction'' when FER \textless{} 0 whilst ``under-prediction''
(FER~\textgreater{}~0) is represented in yellow and red bars. The
examples in Figures \ref{fig2} a and b show the
corresponding mapping functions for total precipitation in 6-h (TP)\textgreater{}7 mm, with
precipitation mainly convective in a, showing an exponential pattern
but mainly stratiform in b, with a more Gaussian shape. We see a
better performance of the model in the stratiform case because
sub-grid variability is much lower. However, in convective
situations, a "good forecast" (white bar, observations within \textpm 25\% of forecast gridbox value) is rare, and overprediction (green) is quite common. Also, red bars are slightly higher
than they are for the stratiform case, which denotes a large
underprediction, which can in turn connect, on occasion, to
flash-floods events that it would have been hard to foresee using the raw model data alone. The other two mapping functions shown demonstrate
the effect of orography.
Figures \ref{fig2} c and d are
for the same basic weather conditions (mainly stratiform precipitation,
4-7 mm in 6-h); but the first one (c) corresponds to flat areas
and the second one (d) to areas with more complex (sub-grid) topography. We can see how
the model performs less well in topographically complex regions, where the
incorrect representation of those topographic details on the model grid (due to resolution limitations) tends to lead to
overestimations (green bars) and significant underestimations (red
bars) of rainfall occurring more often than they do in flat regions. The gridbox mean bias values indicate slightly greater overprediction over mountains (FER=0.85) than over flat areas (FER=0.92). 

\begin{figure}[H] \centering \includegraphics[width=35pc]{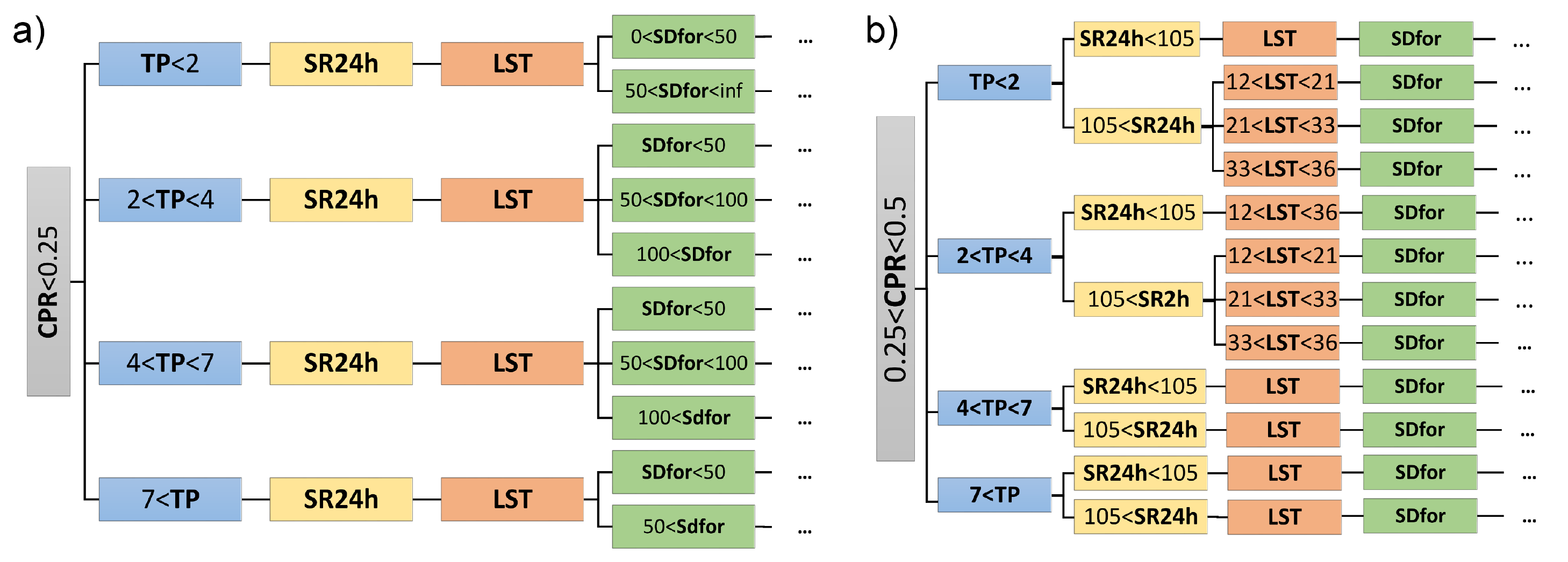} \caption{Two small segments of the final decision tree, from the main branches
(level 1, grey) of (a) CPR\textless{}0.25 and (b) 0.25\textless{}CPR\textless{}0.5. In the
full tree, there are 245 branch terminations or ``leaves'', with 7
levels in total. Each leaf has one mapping function (FER distribution) associated (for examples see Figure 2).
Note the differences between (a), where there is no branching for SR24h and
LST but multiple branches for SDfor; and (b), where SR24h
and LST have branching but SDfor does not.}  \label{fig3}
\end{figure}

To arrive at governing variables for the decision tree, used
to define the gridbox-weather types, the ecPoint philosophy combines
physical reasoning, meteorological experience and the use of case
studies to select variables that have the potential to influence
sub-grid variability and/or gridscale model bias in 6-h precipitation
totals. As in \citet{Pillosu_Hewson2021}, we tried to avoid 'double counting' by not using
two (or more) variables that define much the same model characteristic (e.g. two ways of calculating CAPE), and by not using variables whose influence on rainfall totals is believed to be well captured by the numerical model itself (e.g. humidity mixing ratio). The 6-h
ecPoint-Rainfall calibration was informed by the pre-existing 12-h
product calibration, but was otherwise developed independently. The 6-h weather type definitions
(Figure \ref{fig3}) are currently based on the
following seven governing variables (listed as used in the decision tree, first level first): convective precipitation ratio (CPR), TP,
24-h clear-sky direct solar
radiation at surface (SR24h), local solar time (LST), standard deviation of the
filtered subgrid orography (SDfor), 700 hPa wind speed (V700) and
Convective Available Potential Energy (CAPE). Five of these had been used in the 12-h calibration: CPR, TP, SR24h, V700 and
CAPE \citep{Pillosu_Hewson2021}. 

Variables SR24h, LST and SDfor are all ancillary variables that do not relate to the NWP integrations, but that do all have a bearing on local rainfall outcomes. SDfor provides a numerical representation of sub-grid topographic complexity, whilst SR24h represents the amount of direct shortwave radiation from the sun that would reach the surface of the earth in 24h if skies were clear ($J m^{-2}$, horizontal plane assumed). The new variable SDfor was introduced in recognition of the impact that topography can have on rainfall (and also considering the particular relevance of this aspect to Italy's complex topography). We see clear evidence of impact on Figure \ref{fig2} c and d. Meanwhile LST was introduced to try to account for the known tendency of the IFS to both develop and decay diurnally-driven convection too early in the day \citep{Haiden2018, Bechtold_2014}.

In creating the decision tree, we first decide upon the breakpoints for the first
variable and accordingly create branches to level 2. Then we consider in
turn each of those branches, looking for breakpoints for the second variable. This process is repeated for the third, fourth,
fifth, sixth and seventh variables. For breakpoint selection
for each level, we use a semi-subjective approach (examining each variable individually). Starting near the lower end of a variable's value range we invent a possible breakpoint, create two subsets, test out, increment the possible breakpoint value, and repeat. For each test checks are performed: the first is to ensure adequate subset sample
sizes, whilst the second looks for significant differences between the two FER distributions using the
the two-sample Kolmogorov-Smirnov test \citep{Massey_1951}. A third check involves a subjective analysis of the relative proportion of
over-predictions (green bars in Figure \ref{fig2}),
good forecasts (white bar in Figure \ref{fig2})
 and under-prediction (yellow and red bars in
Figure \ref{fig2}), again looking for some noteworthy differences. If all checks suggest a valid breakpoint, that breakpoint is retained, leaving only cases with higher values to subset when searching for the next possible breakpoint. This process continues iteratively, working down the hierarchy and across through branches, until the whole decision tree has been grown. In practice the procedure involves occasional backstepping for pragmatic reasons. The total number of weather types obtained in this way in the calibration of 6-h ecPoint-Rainfall was 245.

Some parameters were negated for specific tree branches, based on physical reasoning backed up by tests.
For example for CPR\textless{}0.25, SR24h and LST are not considered key
parameters, since direct insolation will not generally have a strong influence on stratiform cloud and rain (note the absence of sub-divisions for these variables on Figure \ref{fig3} a). However, SDfor was considered
relevant for this type of precipitation, as unrepresented topographic
details in the model can strongly affect the stratiform rainfall distribution at
sub-grid level, at least when tropospheric wind speeds, as represented by V700, are non-negligible (note green boxes on Figure \ref{fig3} a). Simply put, there can be precipitation underestimations in upslope areas and overestimations in downslope areas. For
CPR\textgreater{}0.25, convective precipitation evidently
plays a more important role, so
the astronomical parameters can then have a more substantial influence. Therefore, for these decision tree ``branches'', LST and SR24h are utilised (note yellow and orange boxes on Figure \ref{fig3} b) because insolation strongly influences the development and decay of land-based convection. In respect of LST, the aforementioned diurnal cycle biases in convection imply different FER distributions at different times of the day. 
It is important to also highlight that different branches from one variable may not each have the same
number of branches for the next variable, due mainly to different inter-dependence although subset size can also play a limiting role. For example we can see how in Figure \ref{fig3}b larger TP values do not include any division in the LST variable. This was because mapping function differences seen during the breakpoint tests were negligible.

During forecast construction we consider each gridbox around the world, for each time interval, and for each ensemble member, and to each of these we assign one of the 245 weather types. Then in each instance we convert the forecast 6-h rainfall total into 100 equi-probable point rainfall values, using the respective FER mapping function. We then blend together the values across the ensemble, giving us 5100 possible realisations, or ``virtual ensemble members'', of 6-h point rainfall for each gridbox for each forecast time period. Due to storage constraints, these are are saved as percentiles (from 1 to 99). 

\subsection{COSMO ensemble post-processing}

Our "COSMO raw" ensemble output originates from the COSMO-based Italian modelling system,
referred to as COSMO-LAMI and operationally implemented and maintained
by the HydroMeteoClimate Service of the Regional Agency for Prevention, Environment and Energy of Emilia-Romagna (Arpae-SIMC) in collaboration with the Regional Agency for Environmental Protection of Piedmont (Arpa-Piedmont) and  Centro
Operativo per la METeorologia, Operational Centre for Meteorology, from
the Italian Military Air Force (COMET).

The ensemble has been running daily since the end of 2017 at the CINECA
Supercomputer Centre, close to Bologna; it consists of 20 runs of the
COSMO model at 2.2 km resolution, with 65 vertical levels, starting once
a day at 21 UTC and with a maximum forecast lead time of 51 h. The integration
domain covers Italy and part of the surrounding countries (bounding box latitudes:
34.5°N/48.0°N; longitudes: 4.0°E/21.2°E) with a total of 567x701x65
grid points. The grid is rotated lat/lon (0.02°) with a Southern Pole
of rotation (47°S/10°E). COSMO raw receives hourly boundary
conditions from COSMO-ME-EPS, the 7 km ensemble run by COMET over the
Mediterranean area.
Perturbed Initial Conditions are provided by 20 analyses resulting from
the KENDA data assimilation cycle \citep{Schraff_2016}, which is also run
operationally at CINECA. Latent heat nudging is also applied to ensemble
members using the Italian national radar composite.

As previously mentioned, the scale-selective neighbourhood post-processing described in \citet{Dey_2016a,Dey_2016b}, and \citet{Blake_2018}, was adapted and applied to the COSMO raw, and the final post-processed product was simply called "post-processed COSMO". Whilst \citet{Dey_2016a,Dey_2016b} used instantaneous precipitation rates as the target metric, we used 6h precipitation totals as in \citet{Blake_2018}.

Based on a neighbourhood approach, the scales over which COSMO members' 6h totals reach a
specified level of agreement (S, an integer) were calculated at each grid point in the domain, to give a measure of the location-dependent believable scales for an ensemble forecast, for each one of a set of (6h) time intervals in that forecast. In practice S corresponds to the finest (smallest) agreement scale at which the ensemble members become sufficiently similar to provide a valuable, trustworthy forecast. Here S ranges from 0 (good agreement at the model gridscale) to 5 (very poor agreement between ensemble members), which for COSMO respectively correspond to gridlengths of 2.2km to 24.2km. Here we summarize the method used to derive S, in turn, for each grid point P of the model for each 6h period:

1) The difference in precipitation between the ensemble members is evaluated. The comparison involves examining pairs of ensemble members separately, using all possible pairing combinations. For COSMO, which has 20 members in our study, this means 190 different comparisons.

2) If the forecasts overall are quite similar (differences are below a
certain threshold), then the agreement scale at point P corresponds to the
grid-scale of the model itself. If the fields are not that similar, then
a square neighbourhood size = 3 x 3 grid points, centred upon the point
P, is considered.

3) Then spatial precipitation averages, across the 3x3 grid point sets are computed for for all ensemble members, with those values then compared in the same way, to assess similarity.

4) If this time the forecasts are found to be quite similar then the agreement scale is set to 3x3 (i.e. S=1, to denote 1 "ring" of extra gridpoints used). If the fields are not similar enough, then the scale is increased again to give a square 5x5 grid-point neighbourhood.

5) Until the agreement scale has been set, the code will repeat steps similar to 3) and 4), increasing the neighbourhood size each time. At some predefined level, S=5 in this case (i.e. 11x11 gridpoints), computations will stop even if the agreement has
not been reached - this is for computational efficiency reasons. S=5 equates to a box of dimensions \textasciitilde{}24x24km, which in areal terms is about 80\% larger than an ECMWF ensemble gridbox.

6) Finally, the saved agreement scale S (i.e. a neighbourhood size) defines which points' values around grid point P will be used to deliver the probability
distribution for rainfall at P. For example, if the agreement scale is 1, i.e. equating to 3x3 gridpoints, we use 20x3x3 = 180 different values from the 20 ensemble members. These values ranked define the rainfall CDF for point P, and we
conclude by computing from this CDF percentiles 1 to 99 for precipitation at point P (irrespective of how many different values contributed to its CDF). We use the nearest value technique to compute percentiles, so that where S=0, for example, the 95th - 99th percentiles all equal the ensemble maximum. Percentile values are stored at the same horizontal resolution as COSMO raw. 

7) The previous steps are repeated at all grid points (and then for all of the pre-defined time intervals in the ensemble forecast).

With regard to thresholds used at step (2) above, we followed closely the methodology of \cite{Dey_2016b} (their Sections 3.1-3.2), wherein thresholds have a weak dependence on agreement scale. As in that paper we actually found, in some tests, that agreement scales exhibited little sensitivity to whether thresholds varied in this way or were kept constant.

In the context of steps (2) to (5) above, dry weather constitutes a special case. Whilst one might imagine that all members showing no rain at a gridpoint constituted perfect agreement, warranting an agreement scale S=0, in practice in this scenario the algorithm decides that the threshold has not been met (again following \cite{Dey_2016b}), and comparison extends to the 3 x 3 gridpoint squares. The same logic applies at each iteration, such that S=5 is always used if all members show dry across 11 x 11 gridpoint squares. 

Our scale-selective neighbourhood post-processing is in effect used to compensate for there being insufficient limited area model ensemble members. It can retain rainfall signals at the finest scales when there is good agreement between members, as might commonly happen with orographically-forced rainfall, for example. In such instances we are deeming the finer details in the member forecasts to be reliable. At the same time the approach can spread out information from nearby gridboxes when the member signals differ, and are thus deemed less reliable. Cases in point would be a day of scattered showers, or intra-ensemble discrepancies in time and/or space in the arrival of a front. It is important to appreciate that the agreement scale assigned to each gridbox is based only on the similarity of rainfall forecasts within the
ensemble, and is assessed for each forecast period, so other factors such as orography are not directly used here. Equally, we do not attempt any bias corrections. The post-processed COSMO percentiles computed via this procedure, along with those generated by ecPoint, provide input to the subsequent blending activity used to create the merged product.

\begin{figure}[H] \centering \includegraphics[width=35pc]{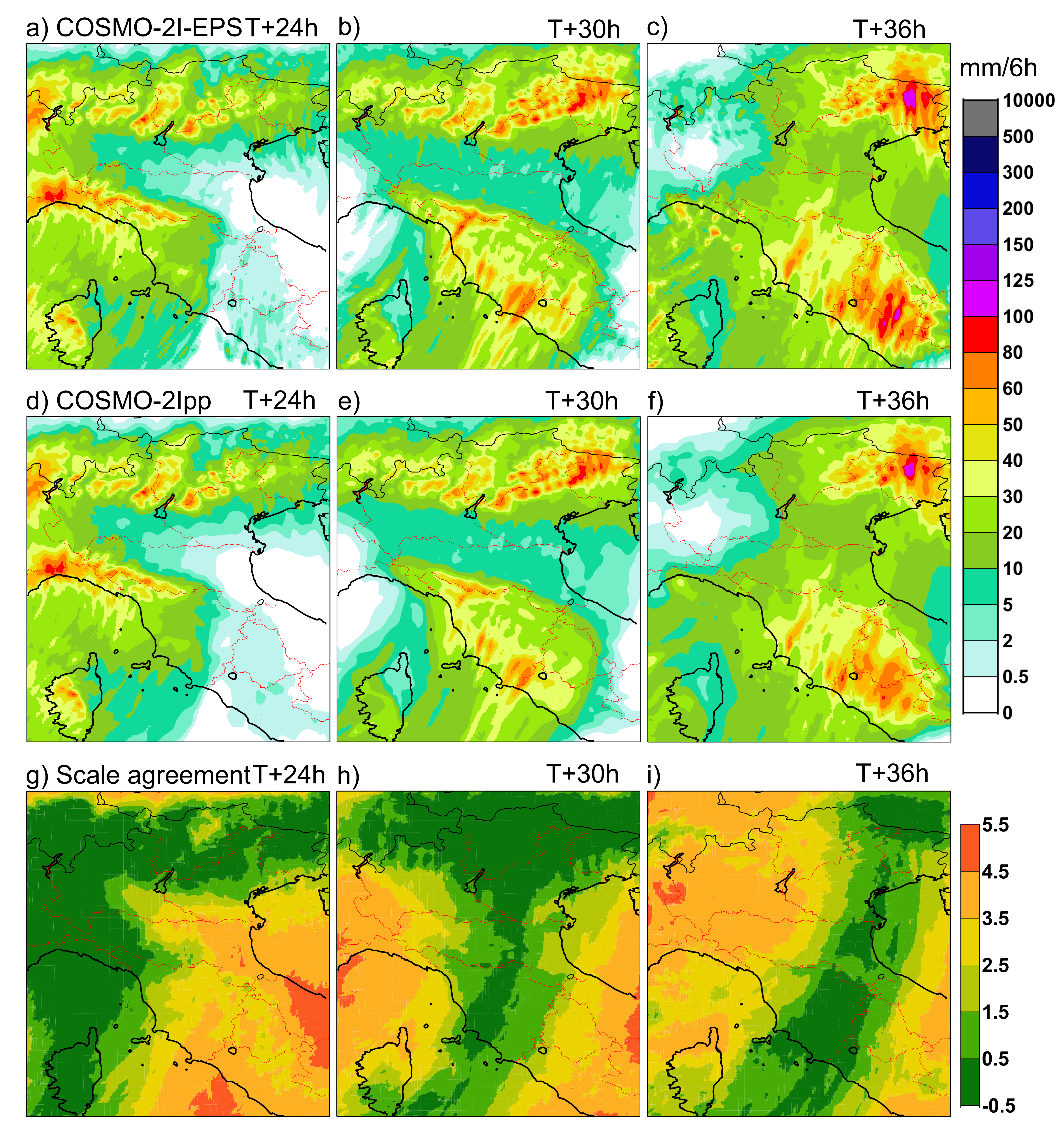} \caption{Examples of the agreement scale and its impact on post-processing of the COSMO ensemble 6h rainfall forecast. All panels show forecasts from a nominal data time of 00UTC on 14th November, valid for (a,d,g) 18UTC 14th to 00UTC 15th (T+24h), (b,e,h) 00 to 06UTC 15th (T+30h) , and (c,f,i) 06 to 12UTC 15th (T+36h). Panels a,b,c show the raw forecast 95th percentile 6h rainfall in mm (=wettest COSMO solution at each point), d,e,f show the post processed 95th percentile, and g,h,i show the diagnosed agreement scale used for post-processing. Agreement scale values are always integers between 0 and 5. Although understanding of 'nominal data time' is not critical for interpretation of this figure the concept is explained in full at the start of Section 3.2.}  \label{fig4}
\end{figure}

As an example of agreement scale usage Figure \ref{fig4} shows, for Northern Italy, diagnosed agreement scale values (g,h,i) along with the corresponding 6-h rainfall 95th percentiles for raw (a,b,c) and post-processed (d,e,f) COSMO, for three consecutive 6h time windows over a night and morning in November 2019. The agreement scales were diagnosed as described above, using all 20 members from one data time.

The integers (S) represented in Figure \ref{fig4}g, h and i are fundamental because they determine how ensemble data is amalgamated during the COSMO post-processing. A value of zero (dark green) denotes maximal agreement in the scheme, and the post-processed forecast is constructed using only values for the said gridbox, across the ensemble. So wherever there is dark green on panels g, h or i, values on the panel immediately above mirror values on the panel above that, implying that local details in ensemble output are being retained because they are consistent enough to be believable. Where the agreement scale is 1 (or more) we use instead values from arrays of 3x3 (or more) adjacent gridboxes to construct the CDF. So details become less and less believable as agreement scale goes up, and are thus removed. In northeastern Italy for example, on panel i, where the agreement scale is mostly 2 (in the southeast corner of Friuli-Veneza Giulia), the post-processed product (panel f) accordingly has a smoother, more spread out look with smaller peaks than the raw output (panel c). On the same panels even stronger "smoothing" is apparent in the Ligurian Sea, where the agreement scale is 3 or 4. There are many papers on the concept of spreading out information, as we are doing here, and such approaches also lie at the heart of the widely used ``fractions skill score'' verification metric \citep{Zacharov_2009, Mittermaier_2010, Skok_2016}.

The apparently eastward-moving band of smaller agreement scales (greens) in this case actually relates to a front, and one can also identify topography-related agreement scale minima - this is all discussed in more detail in Section 3.3 (second case, see also Figure \ref{fig12}). In short, poor agreement between members (large agreement scales) implies that at the model gridscale we have insufficient ensemble members to capture the full range of possible outcomes concomitant with the geographic setting and ongoing synoptic situations. By preferentially spreading information out in such scenarios we should be able to achieve better forecasts for users, and better skill scores in verification, without the large costs involved in running more members. Note however that neither approach can reduce the "intrinsic" predictability of a situation. 

On Figure \ref{fig4}a the large dry patch, remote from the front, corresponds with where S=5 (Figure \ref{fig4}g), in accordance with the aforementioned strategy for dealing with dry outcomes. Figure \ref{fig4}g then also demonstrates how this strategy leads to a smooth, rather than abrupt, spatial transition in agreement scale close to dry areas, with S=4 being diagnosed nearby. Indeed right across the domain, at each of the lead times, S varies smoothly rather than abruptly.

Having relatively smooth spatial variations in agreement scale also tempers spatial variability in grid point value usage in the post-processed COSMO product, thus helping to approximately conserve rainfall "mass". In tests covering different synoptic types the change in domain average precipitation after post-processing was very often between -5 and +10\%, with no clear-cut lead time dependence.

\subsection{Final product: blending together post-processed ensemble outputs}

The "merged product" is the name we give to the final, blended probabilistic 6-h
precipitation forecasts, delivered for Italy and surrounding countries. These forecasts ostensibly go from 6 h to 48 h out (from a nominal 00UTC data time) and have a time resolution of 3-h (i.e. overlapping 6h periods). The blending itself is applied to the ecPoint and COSMO post-processed forecasts (in percentile form), assigning different weights for each of the two components depending on the lead time (the sum of the two weights is always 1). More
weight is given to COSMO at shorter leads, starting with 0.9
at nominal lead time 6 h, with 0.05 of weight decrease every (3 h) time step thereafter until 42 h, after which the decrease is 0.1 at 45 h and 48 h, resulting in an actual weight of 0.1 at 48 h.  The horizontal resolution of the merged product is kept to 2.2 km. Our ``tapered blending'' strategy means that across all lead times the forecast evolution should look relatively smooth to users, with more detail at the shortest leads when the COSMO output is deemed more reliable, due to less synoptic scale spread. The intention was to create a very useful, seamless forecast, with as accurate a representation of forecast uncertainty as the ensembles themselves and our simple "first principles" optimisation approach would allow. Note that \cite{Hemri_2022} provides qualitative support for using a tapered blending approach, and indeed in that study optimal LAM ensemble weights would reach zero around 60 h. Our application of steeper tapering to COSMO weights at long leads was a subjective compromise, to retain seamlessness whilst keeping COSMO weights overall a bit higher. 

Clearly our weighting strategy was not "optimized" using verification. Whilst that would on the one hand be desirable, and is in principle achievable, there would inevitably still be many subjective components, such as metric and threshold selection, and deciding whether or not to employ season-based or weather-type-based dependencies, if data availability allows. At least with our approach there is simplicity and transparency for the user regarding how the merged product was derived; indeed linking the merged product output to the ensemble member inputs in a fully explainable way is also relatively straightforward. Another downside of verification-based optimization is that seamlessness could be intermittently compromised if extra checks were not built in.

In practice the lead times for our operational MISTRAL products do extend all the way to 240 h, but after 48 hours, 
where the COSMO runs end, we just show ecPoint output as is (so for lead times beyond 48 h it is not strictly 'blended' output). In this paper output beyond the 48 h lead will not be discussed further.

The final operational products are in fact created twice a day, blending firstly
6-h ecPoint-Rainfall from the 12 UTC ECMWF ENS runs of the previous day (from lead time T+12 h onwards) with the 21 UTC
COSMO raw outputs of the previous day (from lead time T+3 h onwards), and then later, when the 00 UTC ECMWF ensemble data
becomes available for the new day, that is blended with the same 21 UTC COSMO data, to
deliver an ``update''. However, in this paper, we discuss and evaluate only the
performance of the first blended output as it is the main operational
product. Two different products are operationally produced each day:
probabilities of exceeding specific 6-h total precipitation thresholds
and 6-h total precipitation percentiles. The available thresholds in
the MISTRAL platform are currently 5, 10, 20 and 50 mm for the
probability (of exceedance) product and percentiles 1, 10, 25, 50, 75, 90, 95 and 99
for the second product. Given the way data is stored it would however be simple 
to adjust these specifications if a user requested that. 

ecPoint was specifically designed to improve the
reliability and discrimination ability of the forecast, particularly for
large totals. By blending with the post-processed COSMO output, we aim to improve the forecasts even more, especially in areas of greater topographic complexity, where higher horizontal resolution and, therefore, better definition of the topography is critical. In particular with this blended product we can, in certain situations,
give increased specificity regarding where the most extreme totals are
most likely to occur.

\subsection{The need for HPC infrastructure}

Within the MISTRAL project the Flash Flood Use Case was partly a demonstration of the use of supercomputer architecture to make real-time product delivery tractable; in this instance using the Galileo system (and since June 2021 in Galileo100 system) in the CINECA Supercomputer Centre in Bologna (Italy). The use of HPC resources is crucial for this use case because:

- Considerable computational power and high-quality I/O performance are
necessary for operational production of the COSMO model forecasts. To provide accurate predictions, COSMO has a high-resolution grid and needs to run with
short time steps, both of which elevate the need for computational power. The lower resolution ECMWF ensemble is no less demanding, due to its global coverage, but is run remotely at ECMWF as part of normal operations there.

- Ensemble forecasting requires multiple runs (20 in the case of COSMO) to be performed quickly. This is ordinarily achieved by running members in parallel, which needs a large platform with multiple nodes.

- We need to transform the two ensemble forecasts sets into a much more accurate form, in real-time, using rapid post-processing. In the ecPoint post-processing, for example, we create a new virtual 100-member ensemble for each of the 51 real ensemble members, for each lead time interval. This means we generate 5100 virtual ensemble members for each of 63 time windows, which delivers 7 Tb of output data (for just the rainfall totals) at an intermediate stage. Computational demands are thus very high.

The fundamental requirement above is to deliver forecasts quickly enough for operational use, which means total run times of a few hours at most.

\section{Results}

\subsection{Predictable Scales}

A key user-oriented facet of the merged product is the preservation of high resolution detail (from COSMO) whenever and wherever it is deemed appropriate, via the agreement scales concept outlined in Section 2.3. So here we illustrate typical behaviour of agreement scales over a 1 year period, and highlight some implications.

\begin{figure}[H] \centering \includegraphics[width=35pc]{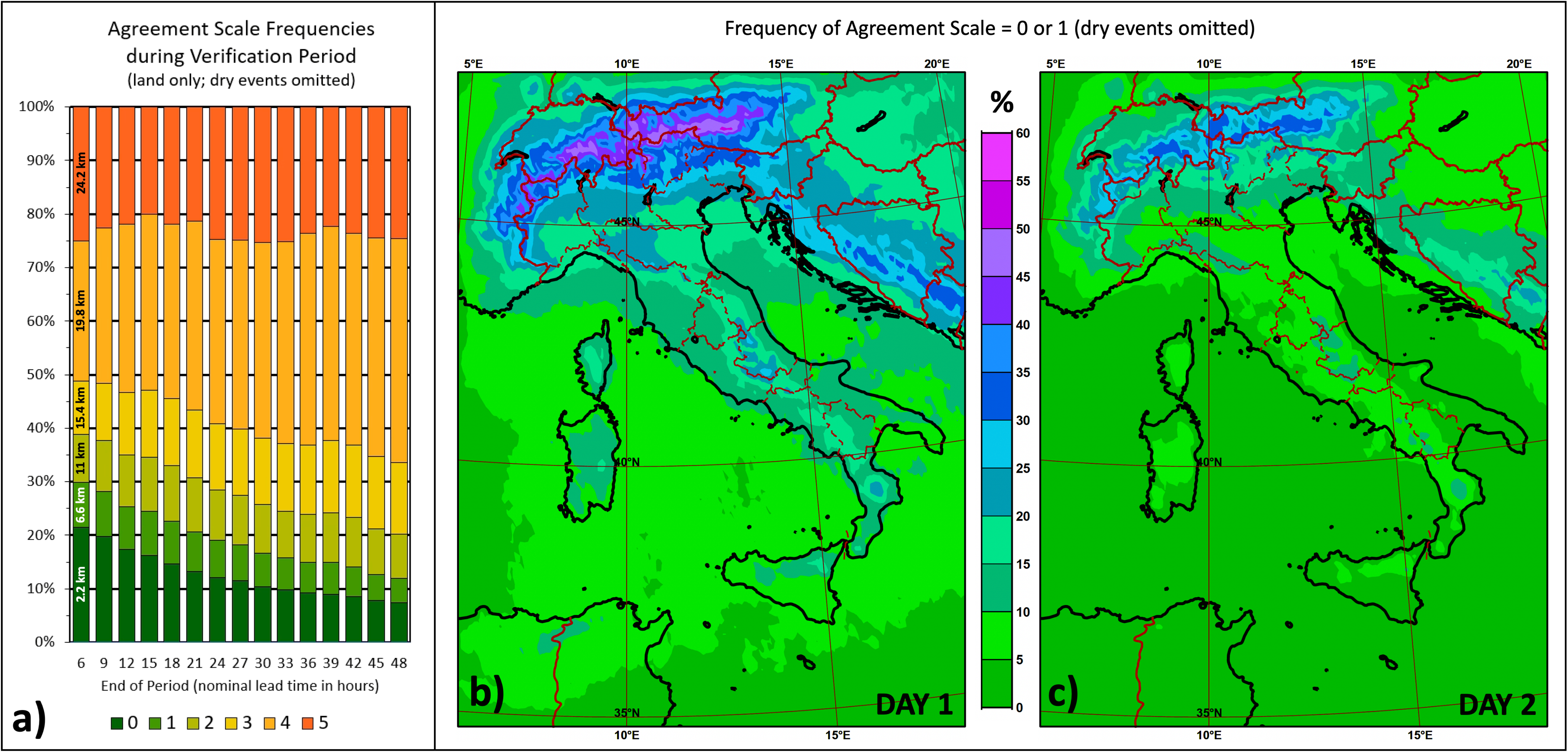} \caption{Agreement scale frequencies (in \%) for post-processed COSMO during the 1 year verification period, for all instances with 99th percentile > 0mm. (a) Value distribution (see legend) versus nominal lead time, for COSMO domain land areas; km values denote effective gridbox dimensions for each agreement scale. (b) Frequency of agreement scale = 0 or 1, over all nominal lead time periods from 0-6h up to 18-24h (day 1). (c) as (b) but for 24-30h up to 42-48h (day 2). Although the understanding of 'nominal lead time' is not critical for interpretation of this figure the concept is explained in full at the start of sub-section 3.2.}  \label{fig6}
\end{figure}

Figure \ref{fig6}a shows that agreement scales tend to grow with lead time in a systematic way. At a nominal lead time of 6h about {50\%} of all non-dry gridpoints have an agreement scale of 3 or less (i.e. equating to gridlengths $\leq$15.4km - Figure \ref{fig6}a), or in other words {50\%} are retaining details at a finer resolution than the current (2022) ECMWF ENS native resolution (18km). By 48h this has reduced to {33\%}. For agreement scales of 1 or less (i.e. 6.6km and 2.2km gridlengths) the corresponding values are \textasciitilde{30\%} and \textasciitilde{12\%}. As the length scales across which COSMO ensemble information is spread out start to surpass the resolution of the ECMWF ENS, ecPoint output (which delivers probabilities for ostensibly similar measurement scales) potentially becomes a competitive and computationally expedient alternative for rainfall prediction, provided its output is well-calibrated across domains with diverse topographies. These results also seem to lend some qualitative support to the tapered blending strategy we developed a priori, with the proviso that verification data needs to also support this.

The diminishing inter-member agreement shown on Figure \ref{fig6}a is intuitively what one would expect, as uncertainties in the synoptic pattern (for example) become ever greater with lead time. Interestingly \citet{Dey_2016b} found no systematic changes in agreement scales with lead time (they examined up to 36h). However we would ascribe that to their use of a summer period, and precipitation rate snapshots as the target variable (rather than totals); presumably convective cell positioning at short leads was already highly uncertain. In our case precipitation total stripes, arising from cell movement, would be more likely to overlap at shorter leads, giving intrinsically smaller agreement scales then.

Another interpretation of agreement scale growth is that it may reflect a need for more and more members, as lead times extend, to usefully span all possible outcomes at fine scale, even when using a post-processing technique like this. This is also what one would intuitively expect. Such a solution would of course have cost implications.

So where is most value likely to be added by post-processed COSMO? Figures \ref{fig6}b and c clearly demonstrate that over mountainous areas there is systematically better inter-member agreement, implying less capacity (and in principal less need) to adjust there to 'add value'. In other words there tends to be less lateral information spread through COSMO post-processing in those regions,commensurate with what one would hope and expect from orographic forcing arguments. On day 1, agreement scales of 0 or 1 are diagnosed as often as 50\% of the time over the most mountainous parts of the Alps, whilst mountain chains such as the Apennines also exhibit higher frequencies than flatter areas nearby. Similar to this, plots in \citet{Dey_2016b} show systematically better inter-member agreement over the more mountainous northern and western parts of the UK. In our case the frequencies diminish everywhere on day 2 (Figure \ref{fig6}c), and whilst a map of day 2 : day 1 frequency ratios is rather noisy (not shown) there is evidence that agreement levels hold up slightly better in mountainous areas. For example over plains north of Bologna values drop from about 18\% on day 1 to 8\% on day 2, whilst in Alpine terrain in the northeastern corner of Trentino-Alto Adige they drop from {45\%} to just 30\%. 

Even if our post-processing is much more 'active' over flatter areas, particularly at longer leads, one should not forget that the targetted retention of more detail over mountains is nonetheless likely to be a key contributor to any general COSMO post-processing success. Indeed \citet{Blake_2018} highlighted that the benefits of the variable neighbourhood approach over the USA, in terms of discrimination and reliability metrics, and relative to a fixed neighbourhood method, were most notable in "the West", which they attributed to that region's greater topographic complexity.

The one year skill of our post-processed COSMO output, and of the other MISTRAL forecast components, is explored in the next sub-section.

\subsection{Verification}

Verification of 1 year of retrospective 6-h rainfall forecasts was carried out, using as truth rain gauge observations from
both standard SYNOP reports and specialised high-density datasets from
Italy and some surrounding countries (see coverage example on Figure \ref{fig1}). Five probabilistic forecast "components" were assessed: the two raw ensembles (IFS ENS raw, COSMO raw), the differently post-processed versions of these (ecPoint-Rainfall,
post-processed COSMO) and the merged version of these two post-processed products (merged product). The verification region coincides with COSMO raw coverage (Italy and surrounding areas, Figure \ref{fig1}), whilst the verification period runs from 1 February 2019 to 31 January 2020 (in January 2019,
COSMO raw was not yet fully operational which meant some days were missing). All verification scores are computed for 6-h non-overlapped periods (for observation times 00, 06, 12 and 18 UTC), for (nominal) lead times T+6 h up to T+48 h. We say here "nominal" because a "T+6 h" nominal lead time (for example) means that for the two COSMO product sets the actual lead time is T+9 h (because the formal initialisation time is 21UTC the day before) whilst for the two ENS product sets it is actually T+18 h (because we are referencing the previous day's 12UTC run, as indicated above). Naturally for the blended product it is a mix. This approach to nomenclature was taken to avoid over-complicating figures and text, but it is something that clearly needs to be borne in mind when interpreting results. The same nomenclature was also used on Figures \ref{fig4} and \ref{fig6}. For the observation periods ending 03, 09, 15 and 21UTC a much reduced observational coverage prevented meaningful verification from being performed, even though (overlapping) products were created for those times.

The reliability component of the Brier Score (BSrel: \citet{Brier_1950} and \citet{Wilks_2011})
was used here to evaluate the reliability of the forecasts. In our case the Brier Score has its binary event form, and
represents the mean-squared error of the probability forecasts, over the
verification sample, for a specific threshold of 6-h accumulated precipitation. Meanwhile, to
evaluate the capacity to discriminate events, we used Area under
the Relative Operating Characteristic curve (AROC; \citet{Richardson_2003}). Three different
precipitation thresholds are initially considered:~\(\ge\)0.2,~\(\ge\)10, and~\(\ge\)30mm/6h. The latter value is quite extreme for most of Italy, and may signify a flash flood risk (depending on other factors). There were about 10000 reports of >30mm/6h during the year. As the COSMO raw domain is quite small, the sample size for thresholds larger than 30mm was too limited to provide robust verification. Following the same verification procedures as \citet{Pillosu_Hewson2021} and according to
\citet{Hamill_2006} recommendations, we added AROC scores for climatological probabilities, based on all available 18-year observation series, to show a ``zero-skill'' baseline (having first discretised to probability intervals of 1\% to mirror ecPoint, post-processed COSMO and merged product discretisation). Conditional verification scores were also computed for different seasons, and also for different SDfor ranges (to divide up according to underlying topographic complexity). For these classes we used an upper precipitation threshold of 20 mm/6h as case counts naturally diminish following subsetting. 
Confidence intervals associated with BSrel and AROC are obtained using 1000-block bootstrap
samples, using the stationary bootstrap scheme with mean block length according to \citet{Politis_Romano1994} and following the same approach as \citet{Pillosu_Hewson2021}. In our verification plots BSrel uses the raw ECMWF ENS as a reference point (i.e. we subtract component performance from that of ENS raw, replicating the reference-run plotting approach of \cite{Baran_2019} and others), so bootstrapping
is calculated from these BSrel differences instead of with absolute values. We also show the actual reliability of the raw ENS for reference (as dashed lines). Including this enables one to also see the degree by which the other systems improve (or degrade): evidently if a forecast system had perfect reliability its solid line would overlay the dashed line.

\begin{figure}[H] \centering \includegraphics[width=35pc]{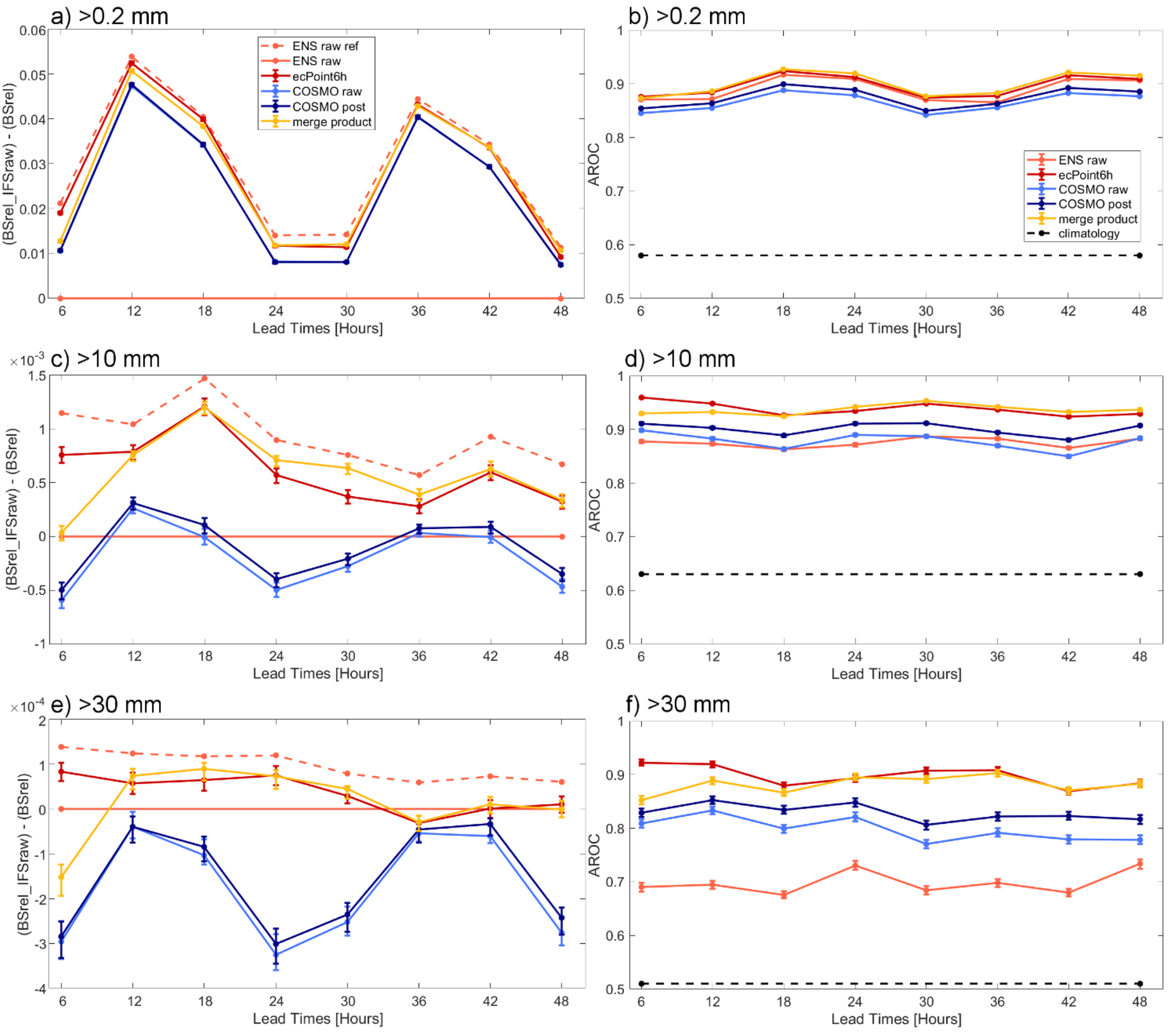} \caption{Forecast verification scores for 6-h rainfall over a 1-year period, using gauge observations, for an extended Italian domain, for nominal lead time intervals of T+0-6h to T+42-48h; x-axis values denote interval end point. Lines denote ENS raw (orange), ecPoint (red), raw COSMO (light blue),
post-processed COSMO (navy), merged product (yellow). Left column (a,c,e) displays BSrel metrics, with dashed orange line showing absolute values of BSrel for ENS raw, and solid lines differences relative to this for the other forecast systems (i.e. ENS raw minus other system). So difference>0 means the other system is more reliable, with that system rated fully reliable wherever its solid line reaches the "cap" provided by the dashed line. Right column (b,d,f) shows AROC as a measure of discrimination ability (with a climatology ``baseline'' shown dashed black); larger is better, upper limit is 1. Each row is for a different precipitation threshold: (a,b) ~\(\ge\)0.2 , (c,d) ~\(\ge\)10, (e,f) ~\(\ge\)30mm/6h. Error bars on all panels show 95\% confidence intervals from bootstrapping.}  \label{fig5}
\end{figure}

Figure \ref{fig5} shows BSrel differences versus ENS raw
(left column), and AROC (right column) for different 6-h precipitation thresholds. First note how
for ~\(\ge\)0.2 mm there is a clear-cut diurnal cycle in ENS raw reliability. All other forecast systems show much improved reliability at all lead times for this threshold, with any residual diurnal cycles in those systems being, in relative terms, negligible (Figure \ref{fig5}a). There is also a diurnal cycle in AROC for this 0.2mm threshold (Figure \ref{fig5}b), but this is seemingly less pronounced and is there in all components. In relative terms, ENS raw is least reliable by day (06-18UTC), perhaps because showery activity, which probably peaks then, tends to lead, in a given ENS gridbox, to some sites recording rain when others nearby do not. ecPoint and the higher resolution ensemble can cope with this, the raw ENS cannot. 

Diurnal cycles in reliability for the raw ENS are not so evident for the higher
thresholds ~\(\ge\)10 and ~\(\ge\)30 mm
(Figure \ref{fig5}c, e). These spurious cycles are again addressed by ecPoint-Rainfall and by the merged product, although as the threshold increases the improvement imparted by the post-processing becomes less emphatic. Meanwhile COSMO products have their own diurnal-cycle reliability errors for the higher thresholds, which are even worse than we see in raw ENS. This could be in part because we do not apply any bias correction to raw COSMO or post-processed COSMO products, which seemingly have particular reliability issues in representing heavier evening and night-time rainfall (18-06UTC). One can say that ecPoint-Rainfall (red line) just about has the best overall performance in terms of reliability and discrimination for the 10mm and 30mm thresholds. The merged product is the best product for one or both metrics some of the time, notably when it can exploit better performance of post-processed COSMO. However, for very short lead times (06 h and 12 h), the merged product can be much worse than ecPoint-Rainfall for both metrics, so the high weight then afforded to COSMO raw in the merging seems inappropriate. By lowering this weight the merged product would have undoubtedly performed better, and this should probably be adopted in a future version. 

The curious short-lead behaviour in COSMO output on Figures \ref{fig5}c,e was briefly investigated by computing, for all leads, annual domain-average over-land 6h rainfall totals. These totals exhibited a decaying downward trend with lead time (superimposed on a diurnal cycle), with 10\% more rain for nominal lead times 00-06UTC on day 1 (T+0-6h) than on day 2 (T+24-30h), reducing to 3\% more for 18-24UTC. This suggests that COSMO forecasts may have had excess rainfall at short leads, degrading reliability then, although in proposing this explanation we were not able to directly compare with observed equivalents, due to inhomogeneous data coverage. Another curious feature of all the reliability plots is that ENS raw reliability (dashed lines) also tends to improve with lead time (discounting diurnal trends). This is qualitatively similar to global verification for 12h rainfall in \cite{Pillosu_Hewson2021}, in which reliability improves up to day 5 then plateaus, and may signify either a "slow spin up" issue and/or insufficient spread. In turn such trends may relate to ECMWF's intention to optimise ENS performance, in terms of spread-skill relationships and other metrics, in the medium range, via tailored singular vectors \citep{Palmer_2019}, although conversely note that the stochastic perturbations also used to create spread are not themselves targetting medium range performance \citep{Leutbecher_2017}.

COSMO post (dark blue line) shows a better resolution and reliability than raw COSMO (light blue line) in all instances (excepting 0.2mm reliability which looks identical). An implication of COSMO agreement scales increasing with lead time (Figure \ref{fig6}a) is that more value can in principal be added by post-processing later on, such that the rate of skill degradation with lead time is reduced. We see signs of this happening on Figure \ref{fig5} and indeed on most other verification plots presented in this section (on Figures \ref{fig8} and \ref{fig10}), most notably for AROC. Specifically the gap between light blue and dark blue curves tends to increase with lead time.

A final point to note here is that, for all thresholds and all lead times, ecPoint-Rainfall has better discrimination ability than both the corresponding ENS raw and the post-processed COSMO forecasts.

\begin{figure}[H] \centering \includegraphics[width=35pc]{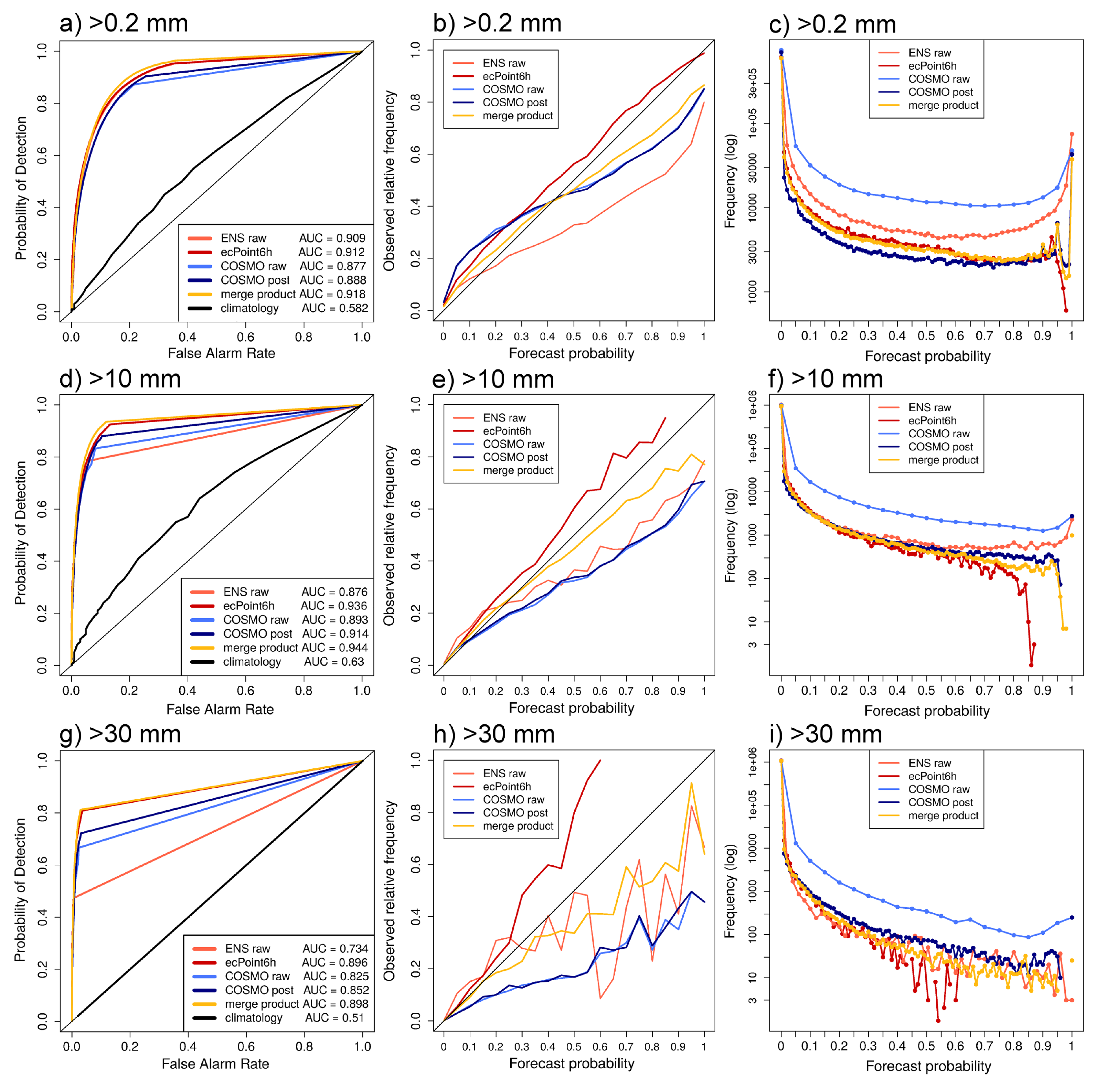} \caption{Forecast verification scores for 6-h rainfall for the nominal 18-24h lead = 18-24UTC, over a 1-year period, using gauge observations for an extended Italian domain as on Fig 4. Line colours also as on Figure 4. Rows signify precipitation thresholds:~\(\ge\)0.2 (a,b,c),~\(\ge\)10 (d,e,f) and~\(\ge\)30mm/6h (g,h,i). Left column (a,d,g) shows ROC curves (with AROC denoted by 'AUC=' in boxes), middle column (b,e,h) reliability diagrams and
right column (c,f,i) sharpness (logarithmic scale on y-axis). For ROC curves and sharpness, probability discretisation reflects the finest available for each forecast system: 1\% for ecPoint-Rainfall, COSMO post,
and merged product; \textasciitilde{2\%} for ENS raw and
5\% for COSMO raw (and 1\% for climatology on ROC curves). Missing points in the profiles on (b,c,e,f,h,i) denote empty classes.}  \label{fig7}
\end{figure}

Figure \ref{fig7} presents ROC (Relative
Operating Characteristic) curves
(Figure \ref{fig7}a, d and g), reliability
diagrams (Figure \ref{fig7}b, e and h), and
sharpness diagrams (Figure \ref{fig7}c, f and i)
for the five forecast products at 24 h lead time. 
As on Figure \ref{fig5} for T+24h the merged product ROC curves exhibit better discrimination ability 
than any of the other 4 components, for all  precipitation thresholds, with forecasts based on climatology performing very poorly, in relative terms. For discrimination for the higher precipitation thresholds (Figure \ref{fig7} d, g), and relative to ENS raw, we also observe clear advantages for both of the COSMO systems, and in the case of COSMO raw this is in spite of its coarser discretisaton. We attribute this to the higher spatial resolution of 2.2 km in both COSMO products, compared to 18 km in ENS raw: probably many cases involve convection or orographic enhancement, both of which can deliver local peaks that are not reflected in the output of current generation global models. At the same time there is the caveat that due to the apparent closeness of the leftmost segments of the ROC curves on Figure \ref{fig7} d, g it is hard to anticipate, a priori, what any of the systems would deliver in terms of ROC area if discretisation were increased (e.g. by running more members).

Whilst the characteristic of ecPoint products having
better reliability and discrimination ability than raw ENS for large totals, previously noted by \citet{Pillosu_Hewson2021}, is supported here, we can also note a reduction in the range of probabilities delivered for such events (Figures \ref{fig7} f and i). No instances have been seen of probabilities >88\% and >62\% for the 10 and 30mm/6h thresholds respectively, for the 18-24h lead time depicted in the 1 year of data. This is not necessarily "wrong" however; note that high probabilities delivered in the two COSMO systems, in ENS raw, and also (to a lesser extent) in the merged product, are all over-confident - on Figures \ref{fig7}b, e and h the respective lines all fall below the diagonal. And high probability case counts are also quite large (Figures \ref{fig7}c, f and i) - e.g. \textasciitilde{2000} for post-processed COSMO for >60\% probability of >30mm/6h. Whilst the larger ecPoint probabilities of >30mm/6h are, conversely, under-confident (line above diagonal), the case count here is relatively low, only \textasciitilde{100} for probabilities \(\ge\){50\%}. Sharpness diagrams plotted in Figure \ref{fig7} use the maximum probability
discretisation that each forecast system can provide, which varies (see caption). Whilst it might look as though there are discrepancies in frequency of occurrence (e.g. COSMO raw seems to exhibit "more of everything") this is just due to the varying discretizations, and also because of relatively large absolute variations in the frequency of zero which on the log scale are hard to discern.

\begin{figure}[H] \centering \includegraphics[width=30pc]{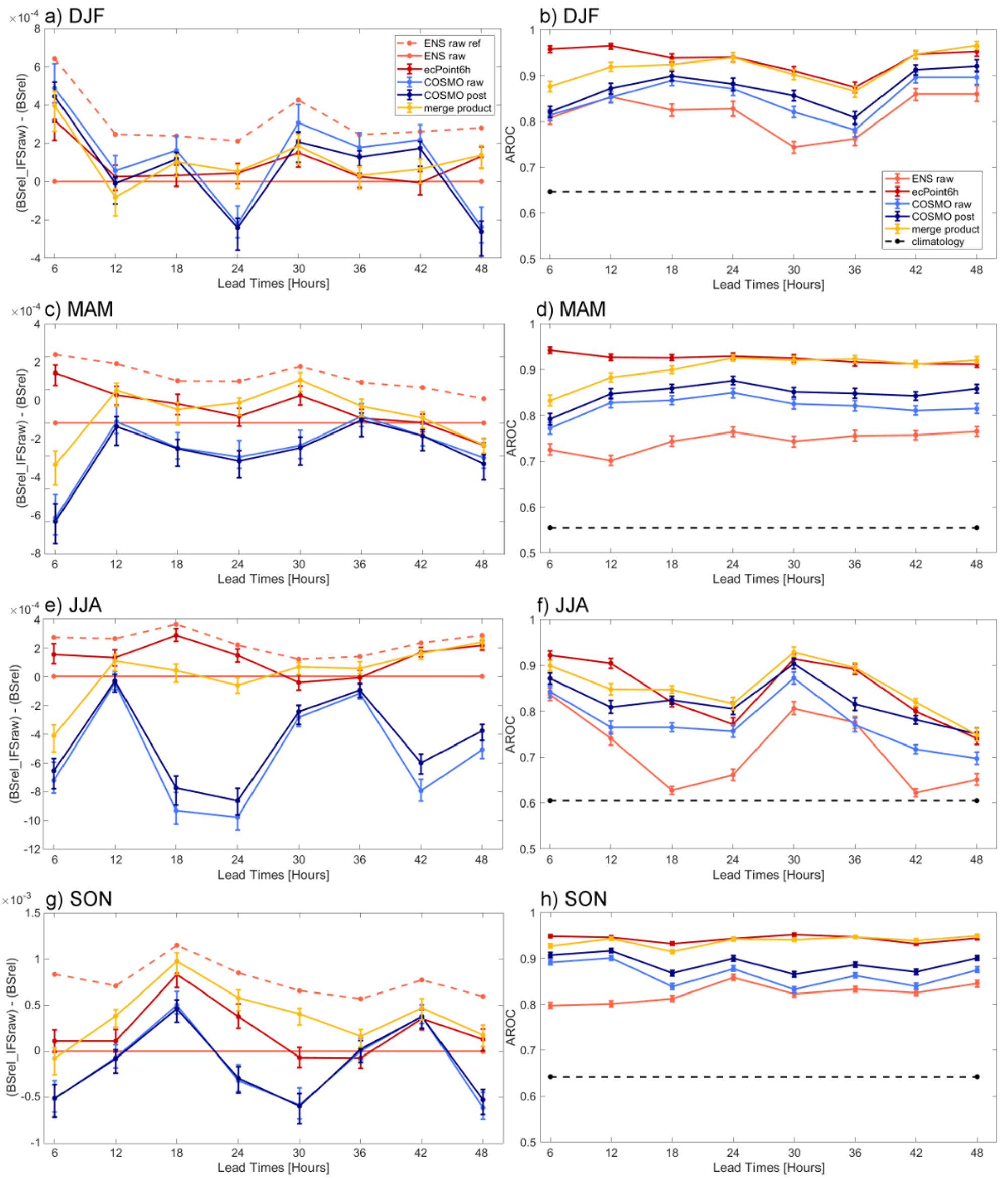} \caption{Forecast verification scores, divided by season, for a 20mm/6h threshold, using gauge observations for an extended Italian domain over 1 year as on Fig 6. Line colours and styles, and x-axis meaning also as on Fig. 5. Left column (a,c,e,g) displays BSrel metrics. Right column (b,d,f,h) shows AROC. Seasons are DJF (a,b), MAM (c,d), JJA (e,f), SON (g,h). All plots include 95\% confidence bars from bootstrapping.}  \label{fig8}
\end{figure}

Figure \ref{fig8} shows a seasonal breakdown of the skill metrics shown on Figure \ref{fig5}, for very wet episodes (here \(\ge\)20 mm rather than \(\ge\)30 mm/6h). The majority of observations of >20mm fell in autumn (\textasciitilde{20000} cases), followed by spring with \textasciitilde{7000} cases, whilst winter and summer saw \textasciitilde{5000} cases each. In summer (JJA) there were two times as many cases in the evening (18-24 UTC) as in any of the other 6-h periods (not shown), which probably reflects a diurnally-driven convective activity peak.

BSrel plots show some differences between seasons (Figures \ref{fig8}a, c, e and
g), but with the highest reliability overall delivered by the merged product. ecPoint output is not far behind, and for T+0-6h = 00-06UTC is the clear winner in three seasons probably because COSMO spin down issues along with the post-processed COSMO's high weight allocation at short leads are contriving to degrade the merged product, as discussed above. Whilst the raw COSMO outputs have relatively strong reliability in winter (DJF) post-processing is, it seems, slightly degrading this: this curious behaviour warrants further investigation. One hypothesis is that the true picture is not emerging because of some erroneous observations in Italy. A high percentage of precipitation in winter is in snow form in the Alps and the Apennines and there may conceivably be problems with the gauge-based heaters that should melt the snow. This could lead to considerable underestimation of precipitation, and after the event sometimes overestimation following in-funnel melting. These aspects may be erroneously nullifying some of the post-processing benefits, although it is also true that skill metrics for the raw ensembles would also be impacted. For ecPoint, a positive design feature of its calibration process is resistance to contamination from occasional erroneous extreme values \citep{Pillosu_Hewson2021}, and in addition its calibration domain is global and not just Italian. Together these aspects elevate the integrity of the ecPoint post-processing for users, but equally limit the scope for error cancellation in the ecPoint verification over Italy. Thus all five products we are discussing are likely to be adversely affected by any errors in the verifying data, although it is difficult to pin down how much in each case.

In summer it is striking how COSMO reliability for wet events is  poor in the afternoon and evening (12-24UTC), on both day 1 and day 2 of the forecast. From Figure \ref{fig7}e,f (for just 18-24UTC) it seems this may be due to over-confidence throughout the probability range, and equivalently to a bias to predict too many wet events. This assertion is supported by reliability diagrams for a time of day when Brier Score reliability is better (06-12UTC = T+12 and T+36, not shown) on which the COSMO lines are closer to the diagonal. So it seems that for COSMO the handling of diurnally driven convection exhibits some pronounced time-of-day related reliability issues.

In terms of discrimination ability for these very wet events, quite similar performance is observed in all the seasons
(Figure \ref{fig8}b, d, f and h): ecPoint and
the merged product are consistently the best, followed by post-processed COSMO, then
raw COSMO, and with raw ENS exhibiting the lowest discrimination ability.
It is also noticeable that the summer season displays a sharp AROC diurnal cycle, particularly for the raw ENS and ecPoint 
(Figure \ref{fig8}f). The ecPoint minima are for 18-24UTC. This is in spite of the ecpoint decision tree having incorporated local solar time dependence to try to mitigate known convective diurnal cycle errors in raw ENS (e.g. Figure \ref{fig3}b), although maybe the picture would have been even worse without such adjustments. Note however that we do not post-process the precipitation field when the model predicts no precipitation, and here this could be a key factor. One could incorporate a new type of dependance by, for example, creating a new ecPoint decision tree branch from the top level, for LST=18-24, that does not multiply forecast precipitation but instead uses additive mathematics based on other relevant model-derived variables, such as CAPE or lightning density, either in the said period or in the preceding 6-hours. Or alternatively one could even use forecast rainfall in the preceding 6 hours in a creative way.

The ecPoint ROC curves, for >20mm/6h, for different times of day for the JJA summer period (not shown) actually exhibit structural differences: relative to the 00-06 and 06-12UTC periods, there is a shift to the right for 18-24UTC, and even more so for 12-18UTC, highlighting a lower innate ecPoint discrimination ability for those later periods. This is broadly consistent with the ROC areas, except that the minimum ecPoint ROC area is actually for 18-24UTC (Figure \ref{fig8}f) and not for 12-18UTC, because for 12-18UTC the points ascend to reach much higher hit rates. In turn this is because 12-18UTC has more usable rainfall signatures in the raw ENS. The higher number of observed cases may also be a contributory factor (in summer the 12-18UTC observation case count for >20mm/6h is $\sim${2200}, versus $\sim${1100} for each of 00-06, 06-12 and 18-24UTC).

During summer evening hours (18-24 UTC), post-processed COSMO is ceutbechrearly better than  ecPoint. The much higher horizontal resolution in COSMO, and particularly a related ability to prolong daytime convection into the evening, which raw ENS lacks, seem to play a critical role here. This clearly benefits the merged product. In the 32 season-lead-time combinations shown the merged product exhibits its biggest gain over ecPoint for JJA T+18-24 (=18-24UTC). The gain would probably also be higher for T+42-48, were it not for the fact that post-processed COSMO weight is then very low. Moreover, the merged product's best season of all, considering all lead times, is clearly summer. Probably, if COSMO reliability were intrinsically better (Figure \ref{fig8}e), or could be made better with some bias correction procedure, the post-processed COSMO performance in summer would be even more striking.

\begin{figure}[H] \centering \includegraphics[width=35pc]{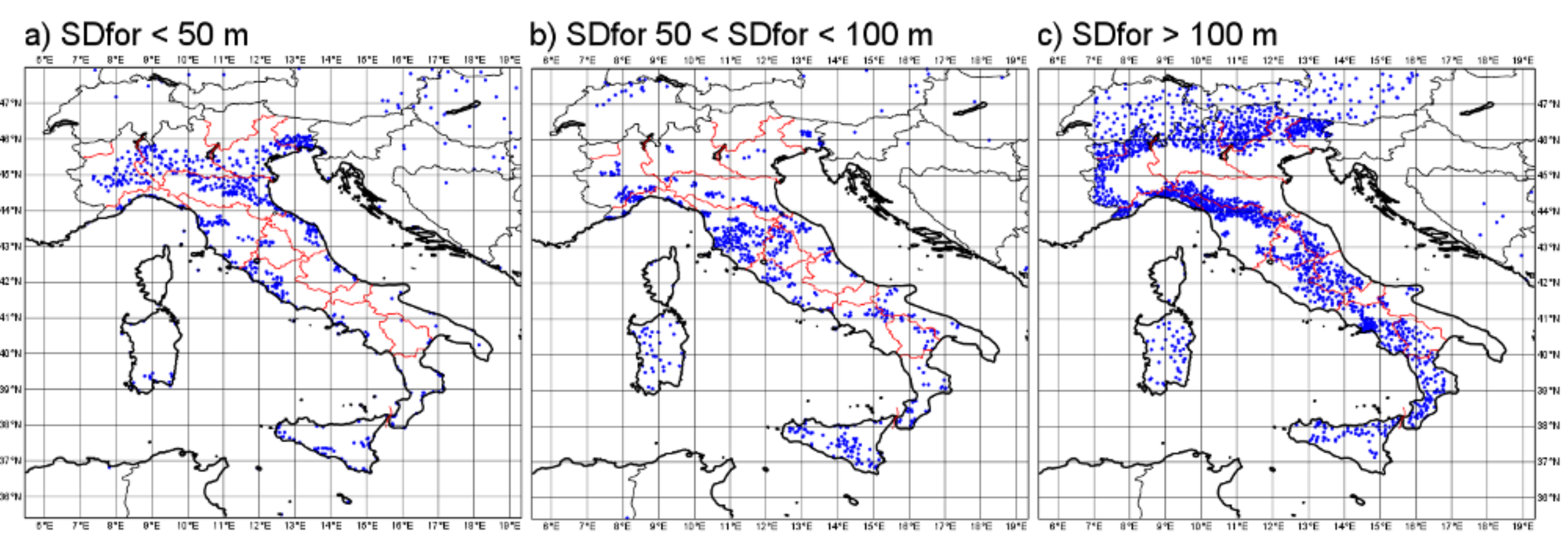} \caption{Observations available for each SDfor range: (a) \textless{}50 m ("plains/low hills"), (b) from
50 to 100 m ("very hilly"), and (c) \textgreater{}100 m ("mountainous")}.  \label{fig9}
\end{figure}

\begin{figure}[H] \centering \includegraphics[width=35pc]{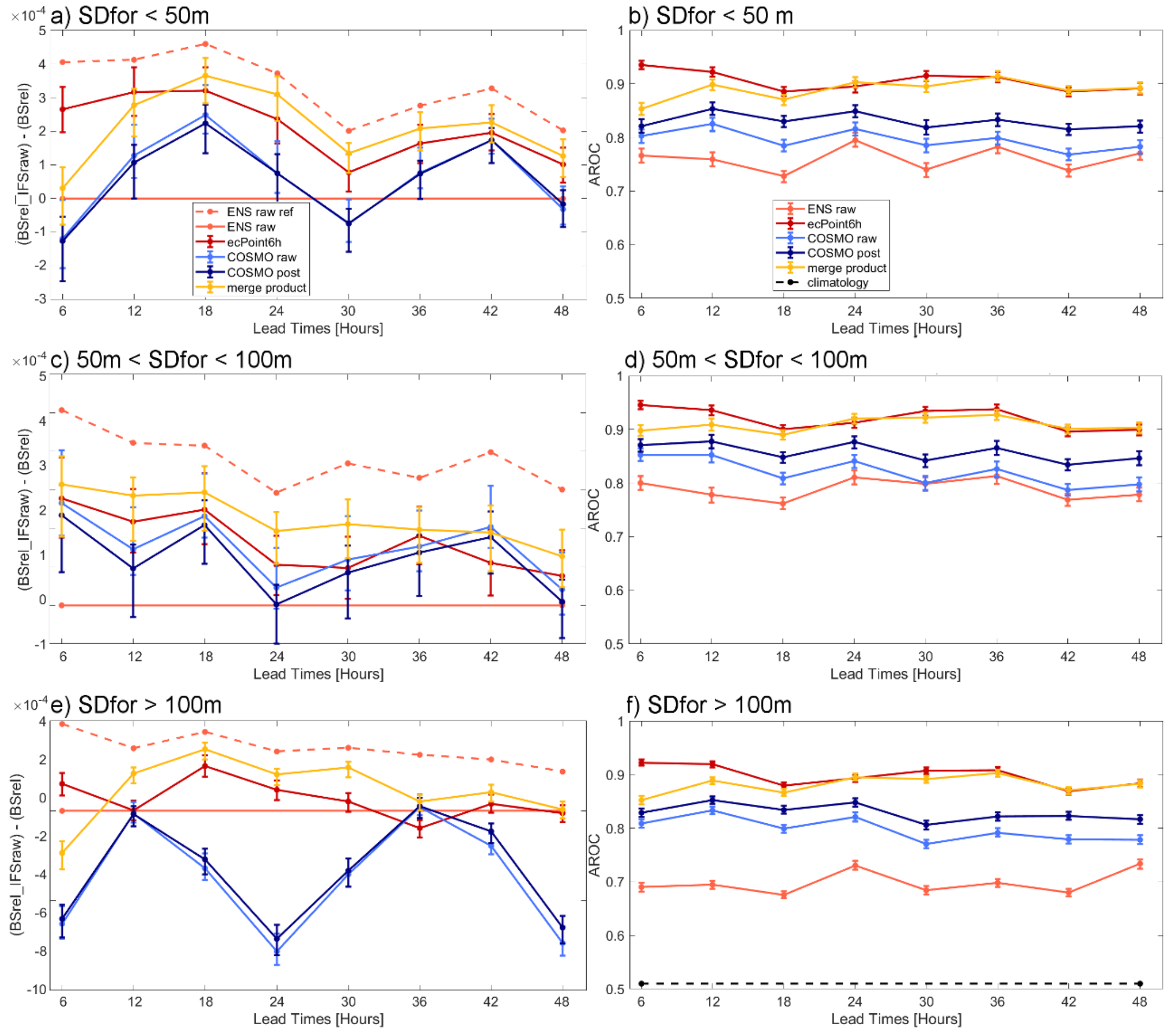} \caption{Forecast verification scores, for categories of topographic complexity, for a 20mm/6h threshold, using gauge observations for an extended Italian domain over 1 year as on Fig. 5. Line colours and styles and x-axis meaning also as on Fig. 5. Left column (a,c,e) displays BSrel metrics. Right column (b,d,f) shows AROC. Topographic complexity is represented by the sub-grid orography variable SDfor, with (a,b) for "plains/low hills" (SDfor<50m), (c,d) for "very hilly" (50-100m) and (e,f) for "mountainous" (>100m). All plots include 95\% confidence bars from bootstrapping.
}  \label{fig10}
\end{figure}

The final verification plots (Figure \ref{fig10}) show performance of the five ensemble
products for different SDfor ranges. As described
above, SDfor is an ENS grid variable which denotes sub-grid topographic complexity, based on a 1 km resolution topographic map and using the standard deviation measure. Higher values naturally correspond to more complex terrain. Figure \ref{fig9} shows the location of observations in three SDfor categories, which we can call "plains/low hills" (a), "very hilly" (b), and "mountainous" (c). Category selection was not arbitrary but corresponds to 
breakpoints found whilst applying the decision tree calibration methodology (albeit with some rounding and set size normalization also applied). As such the breakpoints separate significantly different FER mapping function distributions, for 6h rainfall. In turn this means that we also expect some differences in independent verification for each SDfor range, notably for ecPoint, but also for the merged product which uses ecPoint. Indeed we saw on Figures \ref{fig6}b and c that COSMO post-processing agreement scales also depend on altitude / topographic characteristics, implying reasons to also expect SDfor-related differences in value added by COSMO post-processing. For the different classes the approximate gauge numbers in our domain are 700 (plains/low hills), 900 (very hilly) and 1800 (mountainous). Although not homogeneous these numbers are all still large enough to give a robust annual verification picture, for thresholds up to 20 mm/6h (refer to error bar magnitude on Figure \ref{fig10}).

Figures \ref{fig10}a, c and e for reliability show that ecPoint and the merged product generally perform best, implying some robustness across multiple topographic types, with the merged product best of all. In this regard the capacity of ecPoint to correct the absolute reliability errors of ENS raw (dashed) systematically reduces as topographic complexity increases, from >\textasciitilde60\% improvement for plains/low hills, to <\textasciitilde20\% for mountainous, on average. Strong performance of the ecPoint and merged product components holds also for discrimination ability (Figures \ref{fig10}b, d and f) although here they are on a par. As in verification by season (Figure \ref{fig8}), there are hints of diurnal cycles in the skill metrics, although some of these are hard to interpret. The most clearcut feature in this regard seems to be the AROC minimum for ecPoint output seen for 12 to 24UTC (and by implication a maximum for 00-12UTC). This mirrors what we saw in summer (Figure \ref{fig8}f), but is watered down by inclusion of the other seasons. Notably the signal is there for all SDfor classes.

In mountainous areas meteorological models commonly have difficulty representing precipitation, and this assertion is borne out by the AROC values on Figure \ref{fig10}b,d,f, at least for ENS raw. Whilst the value added by post-processing the COSMO ensemble seems fairly consistent across terrain categories (perhaps contrary to what one might have expected from Figures \ref{fig6}b and c), ecPoint seems, encouragingly, to add most value (to raw ENS) in mountainous regions. It probably reflects the merits of using sub-grid orography as a governing variable in the calibration for mainly large-scale precipitation (Figure \ref{fig2}c,d and Figure \ref{fig3}a). It may also be because there is more discrimination ability to be added in mountainous areas, and because ecPoint post-processing is targetting this effectively via its decision tree subdivisions (even if those have been less effective in improving reliability in mountainous areas). One striking result for the mountainous areas is that the two COSMO products are overall much less reliable than the raw ENS; indeed they exhibit BSrel values that are more than twice as large at some leads (Figure \ref{fig10}e). One might have thought that the COSMO model would have excelled here, especially give that steep-sided mountains and narrow mountain ranges, well below ENS resolution, are a common feature of the Italian landscape. But maybe there are strong biases, impacting reliability, that relate to intrinsic difficulties that even convection-resolving models have in complex terrain. The evening-time drop in reliability of COSMO in mountainous areas is particularly striking, mirroring what we see for summer performance on Figure \ref{fig8}e, and invites further investigation with summer time case studies. Could it be that summer-time evening convection over mountainous regions is a key weak point for COSMO?

\subsection{Case studies}

 We discuss here two case studies of extreme rainfall in Italy, that related locally to flash floods; one was in July, the other in November. The aim is to complement the one-year verification figures, and to also illustrate the spatial characteristics of outputs of our five different forecasting "systems". It was not a given that the products themselves would look physically reasonable and devoid of unwanted artefacts, even if verification is positive, and "believability" of the merged product is of course a key requirement for users. Reassuringly, in our case studies (and others not shown) this aspiration is clearly met. So here we comment on how each forecast component verifies for each case, and to the extent possible link our remarks to verification scores in the previous sub-section.

\begin{figure}[H] \centering \includegraphics[width=30pc]{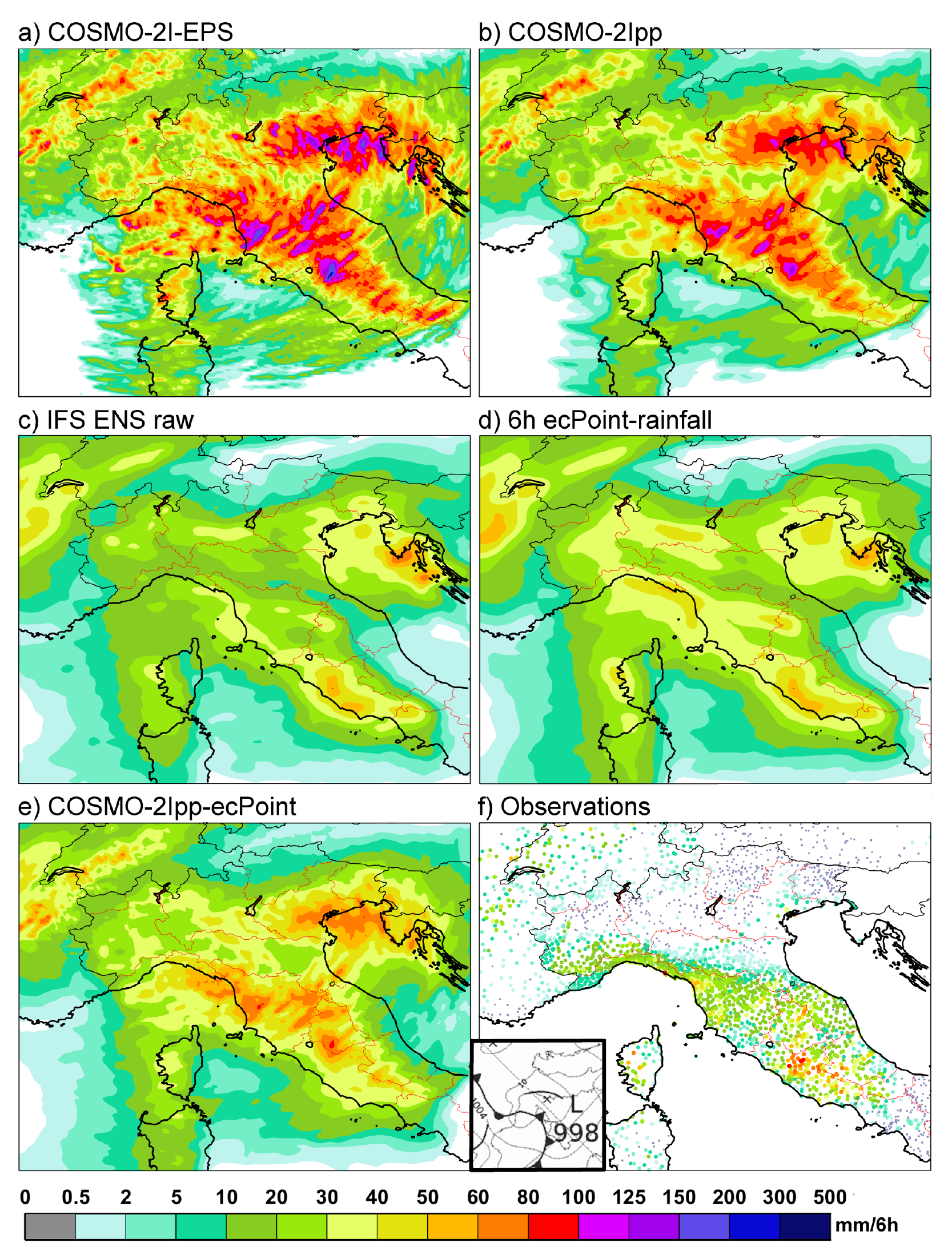} \caption{95th percentiles of 6h rainfall forecasts, for nominal lead time T+24-30 h = 00-06UTC 28th July 2019, from nominal data time 00UTC 27th July, from (a) raw COSMO, (b) post-processed COSMO, (c) raw ENS, (d) ecPoint, (e) merged product. Gauge observations of verifying 6h rainfall are shown in (f); inset shows part of a UK Met Office synoptic analysis chart for 00UTC 28th July, with 4hPa isobar interval.} \label{fig11}
\end{figure}

On 28 July 2019, a shallow cyclone centred in the Ligurian Sea (see inset on
Figure \ref{fig11}f) delivered some extreme
convective rainfall to Italy: in central Italy \textgreater{}40 mm/6h
was widely reported with 80-100 mm/6h locally
(Figure \ref{fig11}f). Flash flood impacts were reported and some rainfall records were broken: https://www.meteoregionelazio.it/2019/07/29/il-forte-maltempo-di-ieri-28-luglio-2019-datifoto-e-video-di-una-giornata-da-ricordare/. 

The 95th percentile forecast from ecPoint (Figure \ref{fig11}d) seems to have performed better in capturing some of the more extreme values than ENS raw (Figure \ref{fig11}c), such as in Liguria and the coastal regions of Tuscany, where ecPoint values are higher, implying some discrimination ability. Whilst some of the extremes appear to not be captured - the highest range shown over Italy is 50-60mm/6h - 99th percentile ecPoint forecast values (not shown) do reach 80-100mm/6h in the wettest parts of the country. In reliable forecasts the 99th percentile should naturally be exceeded 1\% of the time (once in 100 observations), and the 95th percentile 5\% of the time. In fact, the proportion of observations that exceed the 95th percentiles for raw ENS and ecPoint are 23\% and 15\% respectively; for the 99th percentile the corresponding values are 16\% and 3\%. 

The differences between the 95th percentiles for ecPoint (d) and raw COSMO (a) are very striking, even though they are considered to represent similar scales. Whilst we cannot be definitive from one case the raw COSMO rainfall values do look a bit excessive (values of 100-200mm cover quite large areas) compared to the range of values reported (all <100mm), whilst ecPoint looks more reasonable. Post-processing has a marked impact in reducing the majority of the raw COSMO peaks, because of large agreement scales (not shown), although totals on the resulting plot (b) still look too high compared to observations. In this case, the proportion of observations exceeding the 95th percentile for raw COSMO, post-processed COSMO and the merged product are 5\%, 6.7\% and 6\% respectively, whilst for the 99th percentile, the corresponding values are 5\% (same value because of ensemble size), 3.2\% and 1.2\%. So the merged product here seems to be the most useful/reliable overall; it achieves a balance between under-prediction of extremes in ecPoint, and over-prediction in post-processed COSMO. Meanwhile the over proliferation of extreme values at the 99th percentile level in post-processed COSMO may well be hindering it's discrimination ability.

Whilst we cannot do a like-for-like comparison between the stated characteristics for this case and verification scores, because sample size would be too small, there are some useful and meaningful pointers pertaining to the merged product benefits in the data we do have: Figure \ref{fig10}e suggests that COSMO is particularly unreliable for large totals in mountainous areas at night, compared to raw ENS, whilst Figure \ref{fig7}h, together with Figure \ref{fig10}e, suggest this is because of over-prediction. Meanwhile ecPoint generally has much better reliability and discrimination in these circumstances (Figures \ref{fig10}e,f). So the merged product seems to be exploiting, but reducing, the topography-related enhancement delivered by COSMO, whilst benefiting all round from the higher quality but lower-resolution picture delivered by ecPoint.

Success for the merged product also depends on appropriate weighting of ecPoint and post-processed COSMO; here it is 50-50, because of the 30h lead-time and our simple "tapered weighting" approach. The weighting versus lead-time profile is something that could in principal be optimised in some future version, using iterative long-period verification, although that would be computationally challenging.

Over time we have examined outputs of the five different components for many cases. The characteristics we are seeing here are quite typical: for example, post-processed percentiles from the two ensembles tend to resemble their respective raw versions more than they resemble each other, whilst visually the merged product commonly looks like a good compromise, relative to both the input fields and to subsequent observations.

\begin{figure}[H] \centering \includegraphics[width=30pc]{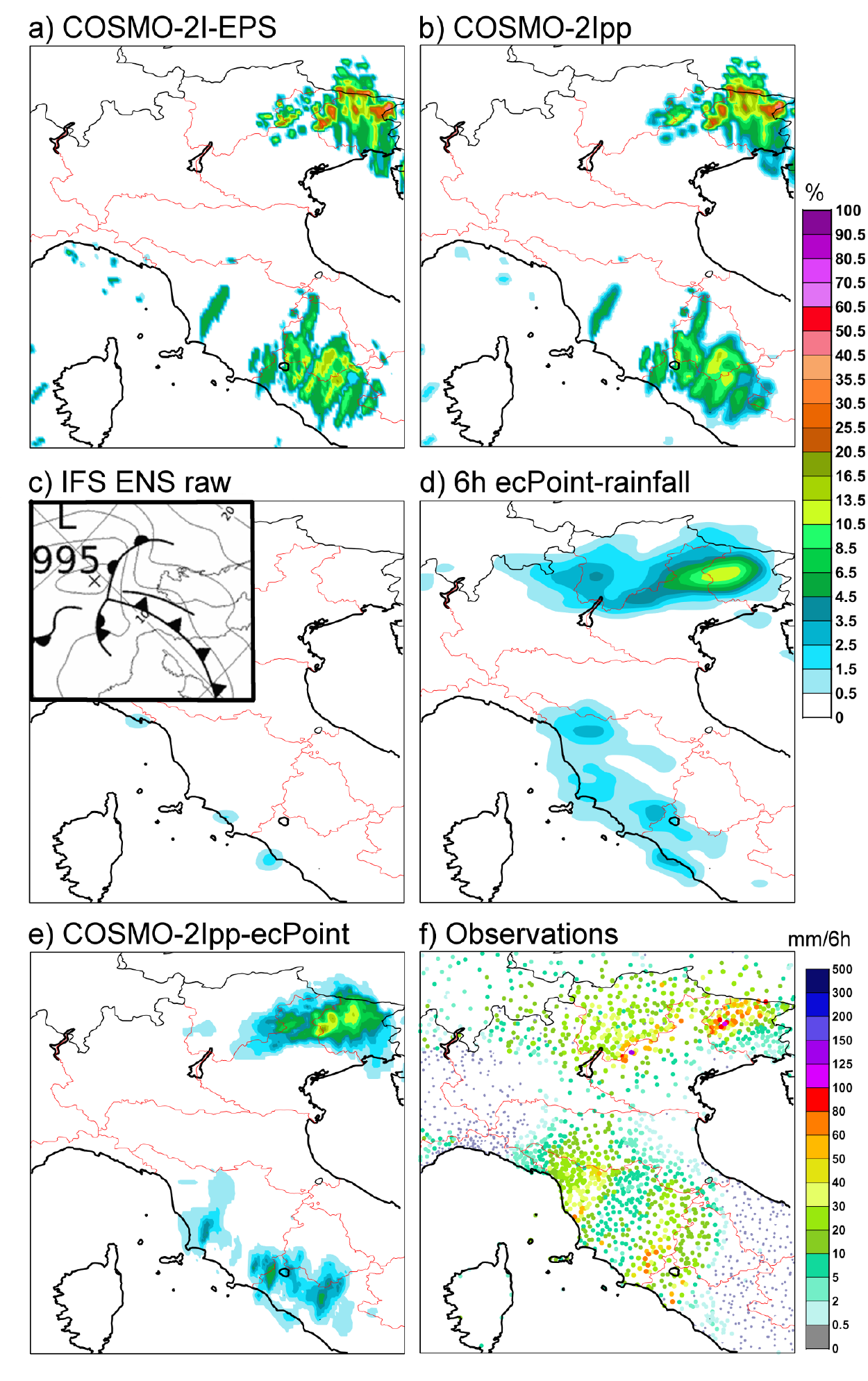} \caption{Forecast probabilities for 6-h precipitation \(\ge\)50 mm, for nominal lead time T+30-36 h = 06-12UTC 15th November 2019 (= rightmost column on Figure 4), from nominal data time 00UTC 14th November, from (a) raw COSMO, (b) post-processed COSMO, (c) raw ENS, (d) ecPoint, (e) merged product. Corresponding observations are shown in (f). Inset on (c) shows part of a UK Met Office synoptic analysis chart for 06UTC 15th November, with 4hPa isobar interval (raw ENS probabilities beneath are all zero). Top legend applies to (a)-(e), bottom legend to (f).
}  \label{fig12}
\end{figure}

The second case (Figure \ref{fig12}) is for 15 November 2019 and corresponds to a very heavy rain event over some northern and central parts of Italy. It is the same case as shown in Figures \ref{fig4} c, f and i. The event was part of a several-day period during which flash floods occurred quite widely: https://messaggeroveneto.gelocal.it/udine/cronaca/2019/11/16/news/nuova-ondata-di-maltempo-in-fvg-allerta-rossa-nel-pordenonese-1.37911793. 
The synoptic pattern over Italy on 15th was cyclonic (inset on Figure \ref{fig12}c), with large-scale 
frontal rainfall seemingly dominant. However inclusion of a warm sector trough suggests a convective component, and the relatively strong warm sector flow also suggests orographic enhancement potential, notably over the Dolomites. In some mountainous areas, for example in Trentino-Alto Adige and northern Lombardy, the existence of large totals alongside negligible totals (Figure \ref{fig12}f) may not be genuine
variability, but may instead by symptomatic of higher-altitude snow not being melted in unheated gauges (the freezing level was of order 1500-2000m).

Panels (a)-(e) on Figure \ref{fig12} show probabilities for precipitation greater than 50mm/6h from the various MISTRAL ensemble components. On the two COSMO plots
(a and b) one clearly sees the modelled impact on rainfall of topographic enhancement, especially in northern Veneto, and Friuli-Venezia Giulia, but also on the southwestern flanks of the Apennine chain where south to southwesterly synoptic-scale flow impinges, in for example Umbria and northeastern Lazio. The high spatial variability in raw probabilities seen in all these areas seems to relate to topographic forcing. For fine-scale signals like these that have intra-ensemble consistency - relating to (e.g.) upslope versus downslope rainfall - we want to preserve rather than "smooth out" the rainfall forecasts. The post-processing algorithm aims to facilitate this using the agreement scale metric (Figure \ref{fig4}g, h and i).

In fact, comparing Figures \ref{fig4}h and i with the synoptic chart (for the midpoint time of 06UTC) on Figure \ref{fig12}c, one can identify an approximately front-parallel agreement scale "trough" (with values of 0 or 1 at its centre) moving eastwards with the front, which in this case is actually the primary driver of ensemble consistency. However there are also lateral topography-related minima superimposed on this band - e.g. in the Apennines along the northern Border of Tuscany, and more generally in the Alps (features which are not inconsistent with the annual average geographical pattern on Figure \ref{fig6}c). Thus the COSMO post-processing picture emerging for this case, as represented on Figures \ref{fig4} c and f, and Figures \ref{fig12} a and b, is quite complex. It involves minimal spreading/smoothing being applied in the zone where the ensemble is confidently indicating that frontal rainfall will prevail in the given time window (e.g. over central and eastern Tuscany), with fine detail retention being further enhanced in upslope/mountainous regions in and beyond that zone (e.g. near the Austrian border of Friuli-Veneza Giulia).

Meanwhile the larger agreement scales (3 or 4) diagnosed in mountainous eastern Umbria and northeastern Lazio exist because frontal timing differences within the ensemble are the dominant factor here. The post-processed COSMO probabilities (Figure \ref{fig12}b) are as expected smoothed out more in these particular regions (in one case with a peak reduced from about 15\% to 7\%). Given the negligible amounts of rain reported (Figure \ref{fig12}f), presumably due to slower frontal progression, such probability reductions can be considered an improvement in the precipitation forecast for this case.

Large precipitation totals (>50mm/6h) were not really directly predicted by any of the ECMWF ensemble members (Figure \ref{fig12}c). Conversely, the ecPoint product (d) had finite probabilities that delineate quite well the areas affected by totals >50mm/6h. The highest probabilities are in Friulia-Venezia Giulia where the observed local frequency of such totals looks greatest, even if the 11-13\% probability forecast (yellow) is arguably not high enough (Figure \ref{fig12}f) from a user perspective. Meanwhile post-processed COSMO (Figure \ref{fig12}b) generally had higher probabilities where observations were largest, although there were a number of "missed events" - most notably in northwestern Veneto where the largest total of all was reported. Thus our semi-subjective assessment concurs with the July case shown earlier: by blending together the two post-processed products we have created a merged product (Figure \ref{fig12}e) that highlights their combined strengths, and that therefore looks more useful than either did in isolation.

The above results can also be cross-referenced with the most appropriate long-period verification data that we have (in section 3.1). Overall that data shows that for high totals, in autumn, for very hilly or mountainous areas, and at a nominal lead time of T+30-36, ecPoint, post-processed COSMO and the merged product are on average equally reliable (Figures \ref{fig5}e, \ref{fig8}g, \ref{fig10}c,e), whilst for discrimination ability the merged product and ecPoint are generally best, with post-processed COSMO significantly less good (Figures \ref{fig5}f, \ref{fig8}h, \ref{fig10}d,f). So the conclusions from our November case look to be fairly typical.

This example also demonstrates a disadvantage of the COSMO post-processing: when (frontal) timing issues contaminate the picture we can end up smoothing out reliable fine-scale details (e.g. upslope and downslope rainfall), that are a characteristic of the wet solutions, for the wrong reasons, such that we end up mirroring a disadvantage - i.e. a lack of geographical specificity - of the lower resolution ensemble. Ultimately this is because as lead times increase the smaller native ensemble size becomes more and more of a limiting factor, calling for increasingly complex post-processing to be employed for the retention of reliable output on fine scales. In our blending system the strategy to counter this, in an imperfect but moderately effective way, is to reduce COSMO weight at longer leads. Perhaps there are lead-time limits beyond which no post-processing method for a finite size convection-resolving ensemble, however sophisticated, could counteract the detrimental impact, on fine-scale rainfall details, of the growth of synoptic scale spread. In which case the optimal low-cost solution, as the length-scale to which S corresponds approaches the ecPoint output grid resolution, could be to just use ecPoint output instead. In the current study and others that relate (\cite{Hemri_2022, Harris_2022}; Cristina Primo-Ramos - personal communication), ecPoint becomes very competitive for lead times greater than 24-48 hours.

\section{Discussion and conclusions}

This paper describes how two innovative, state-of-the-art post-processing approaches have been applied to the output of two different types of ensemble, in order to forecast rainfall in Italy and nearby regions. The final product is created by blending. The overall aim was to deliver much greater skill than can be achieved by using the raw ensembles' output as is, particularly for localised extreme rainfall which is a very common driver of flash flood events. Indeed this initiative was the 'flash flood use case' within the EU-funded MISTRAL project. Whilst the project itself formally ended in January 2021, products continue to be produced in real-time and are accessible on an open-access web platform.

Input data comes from one LAM ensemble (COSMO) and one global ensemble (ECMWF). These ensembles have horizontal resolutions of 2.2 and 18km respectively: in the LAM convection is resolved, in the global model it is parametrised. Post-processing of the two ensembles proceeds in different ways - employing methods commensurate with their resolutions, their capabilities and the lead times they afford. For the LAM, a scale-selective neighbourhood post-processing technique was applied, primarily to try to compensate for there being insufficient ensemble members to capture all the convection-related degrees of freedom. Despite being scale-selective, the technique nonetheless preserves, probabilistically, signals at  2.2km scale. We consider this to be reasonably representative of point (raingauge) scale. For the global ensemble, we use instead the ecPoint downscaling technique. Its main purpose is to activate local-weather-dependant adjustments for gridscale bias and for sub-grid variability, to predict, probabilistically, values as measured by raingauges. Post-processing as described thus puts outputs sourced from the two ensembles onto compatible scales, such that a simple blending step can then be meaningfully performed, preserving fine-scale signals where appropriate. A much more straightforward alternative to the above would be to simply use conservative interpolation to upscale the LAM output to the global model scale before blending, but that would waste the high resolution detail (irrespective of whether it was considered reliable or not) and could not deliver forecasts of high impact localised extremes.

When set alongside the fast-evolving world of artificial intelligence (AI) the scale-selective neighbourhood post-processing we have used could be seen as rudimentary. On the other hand its simplicity is a major strength. Rather than using AI we rely instead on the complexities of the ensemble integrations themselves to define what the 'believable scales' of rainfall prediction are, ultimately for any weather scenarios that may arise. We then encapsulate those in a very simple and interpretable way, as an agreement scale integer between 0 and 5.

In our system blending in the first 48h is simply based on a tapered weighting scheme, with post-processed COSMO output afforded highest weight early on. With this strategy we aimed to best exploit the innate resolution-related advantages of COSMO over ECMWF at short leads. Such advantages are seen for orographic influences, as demonstrated in a one year agreement scale analysis, but also for convection, as reported elsewhere, and other aspects.  Whilst the net reduction in discrimination ability over the 48 hours for raw COSMO may be fairly modest, for larger totals in summer and in autumn, when flash flood frequency over Italy tends to peak, it is more pronounced. And although some of this reduction can be offset with post-processing, ecPoint output is still more competitive at the longer leads. Meanwhile the fraction of COSMO land gridpoints inheriting probabilistic information from neighbourhoods that were smaller than 18km (=ENS and ecPoint output resolution) dropped from \textasciitilde{50\%} at T+6 to \textasciitilde{33\%} at T+48, with km-scale retention dropping away even more rapidly, to \textasciitilde{12\%} by T+48. Such reductions occur because synoptic scale spread growth is rendering the small number of COSMO members increasingly restrictive for longer leads. So for those longer leads the benefits, for rainfall prediction, of investing in costly high resolution runs, versus using ecPoint output alone, reduce, because the two end up largely providing forecasts on similar scales. This provides further support for our strategy to increase ecPoint weights for longer leads. And our approach to make this weight almost one by (nominal) T+48, where COSMO runs end, also ensures that products are seamless across that barrier. Products beyond T+48 are provided by ecPoint alone; these then end at T+240. The focal point in this paper has been the first 48h.

The complex process of post-processing and blending could not have been operationally implemented without HPC resources. These were provided for us by CINECA in Bologna. For the users, the blended point rainfall forecasts are stored as percentiles 1,2,..99, for overlapping 6h periods. Whilst 6h is the shortest period for which there is currently sufficient global coverage of observations for ecPoint calibration, it is also a reasonable compromise from a computational perspective. Although periods shorter than 6h might benefit flash flood applications, that would increase the HPC resource requirement, compromising delivery times to the detriment of users.

Verification for the COSMO area shows that relative to raw COSMO the scale-selective neighbourhood technique generally improves discrimination ability and reliability for different precipitation thresholds, although the differences for reliability are not that significant. Relative to raw ECMWF ENS, ecPoint delivers noteworthy gains in reliability and resolution across multiple thresholds, as reported also by \citet{Pillosu_Hewson2021} in global verification for 12h totals. However ecPoint's reliability gains are bordering on negative for large totals at the longer leads of 36-48h, reflecting a curious improvement in the reliability of the raw ENS then.

Relative to post-processed COSMO, the ecPoint product generally provides better forecasts for all precipitation thresholds examined, although in some circumstances, when considering larger thresholds (e.g. 20-30mm/6h), this is not clearcut. Indeed for large thresholds the advantages of merging the two post-processed outputs are noteworthy for discrimination ability in the summer period, and reliability is generally slightly better for the merged product, for 10 and 30mm/6h, over the year as a whole. Outside of summer time discrimination capabilities for large totals are generally similar for the merged product and ecPoint, except at short leads where ecPoint performs best overall, probably because post-processed COSMO was then afforded too much weight. Spin-down issues seem to have compromised the raw COSMO performance for those short lead times, and the current COSMO post-processing method cannot correct for those. Future releases could address this using e.g. lead-time dependant bias corrections, if appropriate, or more simply by changing the weighting function. 

A particularly strong diurnal cycle in the performance of the different ensemble products was found for summer. Indeed an important result from the summer verification was that the post-processed COSMO improves upon ecPoint in the nominal "evening" period - i.e. 18 to 24 UTC. This is a time when flash floods can often occur, and so is useful to highlight the benefits of blending.

The relatively poor summer time evening performance of ecPoint shows that our attempt to correct for known IFS under-prediction of convective rainfall late in the day, using local solar time as a predictor, has not been that successful. One limiting factor in this regard is undoubtedly the current ecPoint policy to apply no adjustments to forecasts of zero rain. One could in principal develop another decision tree branch, at the top level, for cases of 0mm forecast, trained on what happened when 0mm was forecast, using for example convective rainfall predicted for the previous 6 hours, when local solar time indicated evening, as one predictor. This is another area for future work.

Regarding verification by topographic complexity, ecPoint and the merged product exhibit robust performance for large totals (>20mm/6h) across all types of terrain (found in Italy) in our two verification metrics. Overall these outputs rank as the best two for all lead times, for all topography classes, with the merged product demonstrating better reliability than ecPoint beyond nominal T+6. As expected, our verification also confirms that both COSMO products have much better discrimination of large totals in areas of complex orography than the raw ENS, which is almost certainly due to the innately better representation of complex orography afforded by COSMO's 2.2km spatial resolution. 

Another striking topographic feature is the reduced ecPoint reliability for the 06-12 UTC period in mountainous areas (SDfor > 100 m). The full ecPoint decision tree was designed in such a way that the impacts of both LST and SDfor were never considered in the same branch (for a relevant portion see Figure \ref{fig3}). SDfor was presumed to be particularly relevant for large-scale precipitation (convective component <0.25), whilst for all other classes it was not used, and SR24h and LST were used instead. However, we should perhaps now revisit this, to see if also including SDfor as a governing variable for convective events improves scores, for morning times in particular but others too. As with many ecPoint calibration results this could also provide useful physical pointers for IFS code development.

Analysis of two case studies using plan-view charts helped us to visualise and better understand the impacts of both post-processing techniques, and merging, on the raw forecasts. The merged products looked plausible and were devoid of strange artefacts, satisfying basic user requirements. In general, for the largest (localized) rainfall totals in the two cases, we observed an overestimation in raw COSMO and a significant underestimation in the raw ENS. Post-processed COSMO and ecPoint provided more realistic-looking peaks. Whilst ecPoint also delivered useful guidance on where, in general, the worst affected areas would be, post-processed COSMO was sometimes able to add useful local detail onto that picture. Ultimately users would see a final merged product that looked more credible, more accurate and more useful than any of the input components on their own. Although we have not unequivocally linked particular rainfall measurements to flash flood events, the two episodes examined coincided with times when flash floods were being quite widely reported.

Via the scale-selective neighbourhood technique for COSMO we could see in the case studies where the forecast fine-scale detail tends to be most reliable. As expected, topographically complex areas stand out. In one case a front-parallel agreement scale minimum, centred near to the most likely position of the frontal rainfall, was also very evident. Agreement scale increases either side were symptomatic of uncertain timing of the front. At longer leads, and as frontal timing concurrently becomes even more uncertain, it is logical to expect a general elevation of agreement scales, which is what we saw in our case. In turn this would diminish, with lead time, the innate usefulness of a high resolution ensemble (with a limited number of members) for predicting local details. Out of the LAM ensembles with resolution <3km operational in Europe in December 2021 (see http://srnwp.met.hu/C\_SRNWP\_project/Eumetnet\_List.html), none comprised more than 21 members, and only one system - a 2.2km COSMO ensemble for Switzerland - extended beyond 54h leads. Using differently post-processed output from that ensemble blended with ecPoint output, Hemri et al. (2022) have recently shown (using the continuous ranked probability score on 1 year of data) that value added to what ecPoint alone delivers diminishes with lead time: it was minimal between 60 and 84h leads, and nil beyond 84h. So our inference and their results basically agree.

Looking to the future, this study provides pointers to how real-time numerical prediction of rainfall (and other parameters) could evolve, as LAM ensembles become increasingly widespread, and as global models advance towards having convective-scale resolution. In the short term our blending approach could be adopted for other domains where other LAM ensembles run, perhaps bolstered by verification-based weighting optimisation, and some form of LAM bias correction where necessary. In the longer term, in what will be a new and very different era for global prediction, one can envisage a crunch point arising, where one can afford to run a medium range ensemble with convective-scale resolution, but not the hundreds or thousands of members needed to address both convective and synoptic scale uncertainties beyond the first few days. One could accept the jumpy medium range output that would arise with fewer members, though users may be dissatisfied (note that even 18km resolution ensembles can be jumpy for extreme events - see Figure 3 in \cite{Pillosu_Hewson2021}). Or one could conceivably invent new post-processing approaches - simple or complex or a mix - to quell jumpiness and recover skill, but currently the route to that is not clear. On the other hand, and this is a very open question, it may be that high resolution global ensembles do advance the predictability frontiers for synoptic scales, by capturing important upscale effects, which would make longer lead usage vital even with the member number limitations. But if this proves to not be the case then another option would be to terminate convective scale global ensembles at some intermediate lead time, say 4 or 5 days, whilst also running at lower resolution to much longer leads, and use post-processing and blending techniques like those described here to achieve seamless jump-free output, not just locally but globally. In the end it will be a question of how to best utilise, for users, the expanding and evolving HPC resources, focussing not just on rainfall but also of course on other impact-defining parameters.

\section{Acknowledgements}\label{acknowledgements}

The authors gratefully acknowledge financial support from the European
Union Connecting Europe Facility (CEF) -- Telecommunication Sector
Programme for the MISTRAL -- Meteo Italian SupercompuTing poRtAL
project (Grant Agreement No. INEA/CEF/ICT/A2017/1567101).

\bibliography{biblio}
\end{document}